\documentclass[useAMS,usenatbib,usegraphicx]{mn2e}
\usepackage{color}
\definecolor{AliceBlue}{rgb}{0.94,0.97,1.00}
\definecolor{AntiqueWhite1}{rgb}{1.00,0.94,0.86}
\definecolor{AntiqueWhite2}{rgb}{0.93,0.87,0.80}
\definecolor{AntiqueWhite3}{rgb}{0.80,0.75,0.69}
\definecolor{AntiqueWhite4}{rgb}{0.55,0.51,0.47}
\definecolor{AntiqueWhite}{rgb}{0.98,0.92,0.84}
\definecolor{BlanchedAlmond}{rgb}{1.00,0.92,0.80}
\definecolor{BlueViolet}{rgb}{0.54,0.17,0.89}
\definecolor{CadetBlue1}{rgb}{0.60,0.96,1.00}
\definecolor{CadetBlue2}{rgb}{0.56,0.90,0.93}
\definecolor{CadetBlue3}{rgb}{0.48,0.77,0.80}
\definecolor{CadetBlue4}{rgb}{0.33,0.53,0.55}
\definecolor{CadetBlue}{rgb}{0.37,0.62,0.63}
\definecolor{CornflowerBlue}{rgb}{0.39,0.58,0.93}
\definecolor{DarkBlue}{rgb}{0.00,0.00,0.55}
\definecolor{DarkCyan}{rgb}{0.00,0.55,0.55}
\definecolor{DarkGoldenrod1}{rgb}{1.00,0.73,0.06}
\definecolor{DarkGoldenrod2}{rgb}{0.93,0.68,0.05}
\definecolor{DarkGoldenrod3}{rgb}{0.80,0.58,0.05}
\definecolor{DarkGoldenrod4}{rgb}{0.55,0.40,0.03}
\definecolor{DarkGoldenrod}{rgb}{0.72,0.53,0.04}
\definecolor{DarkGray}{rgb}{0.66,0.66,0.66}
\definecolor{DarkGreen}{rgb}{0.00,0.39,0.00}
\definecolor{DarkGrey}{rgb}{0.66,0.66,0.66}
\definecolor{DarkKhaki}{rgb}{0.74,0.72,0.42}
\definecolor{DarkMagenta}{rgb}{0.55,0.00,0.55}
\definecolor{DarkOliveGreen1}{rgb}{0.79,1.00,0.44}
\definecolor{DarkOliveGreen2}{rgb}{0.74,0.93,0.41}
\definecolor{DarkOliveGreen3}{rgb}{0.64,0.80,0.35}
\definecolor{DarkOliveGreen4}{rgb}{0.43,0.55,0.24}
\definecolor{DarkOliveGreen}{rgb}{0.33,0.42,0.18}
\definecolor{DarkOrange1}{rgb}{1.00,0.50,0.00}
\definecolor{DarkOrange2}{rgb}{0.93,0.46,0.00}
\definecolor{DarkOrange3}{rgb}{0.80,0.40,0.00}
\definecolor{DarkOrange4}{rgb}{0.55,0.27,0.00}
\definecolor{DarkOrange}{rgb}{1.00,0.55,0.00}
\definecolor{DarkOrchid1}{rgb}{0.75,0.24,1.00}
\definecolor{DarkOrchid2}{rgb}{0.70,0.23,0.93}
\definecolor{DarkOrchid3}{rgb}{0.60,0.20,0.80}
\definecolor{DarkOrchid4}{rgb}{0.41,0.13,0.55}
\definecolor{DarkOrchid}{rgb}{0.60,0.20,0.80}
\definecolor{DarkRed}{rgb}{0.55,0.00,0.00}
\definecolor{DarkSalmon}{rgb}{0.91,0.59,0.48}
\definecolor{DarkSeaGreen1}{rgb}{0.76,1.00,0.76}
\definecolor{DarkSeaGreen2}{rgb}{0.71,0.93,0.71}
\definecolor{DarkSeaGreen3}{rgb}{0.61,0.80,0.61}
\definecolor{DarkSeaGreen4}{rgb}{0.41,0.55,0.41}
\definecolor{DarkSeaGreen}{rgb}{0.56,0.74,0.56}
\definecolor{DarkSlateBlue}{rgb}{0.28,0.24,0.55}
\definecolor{DarkSlateGray1}{rgb}{0.59,1.00,1.00}
\definecolor{DarkSlateGray2}{rgb}{0.55,0.93,0.93}
\definecolor{DarkSlateGray3}{rgb}{0.47,0.80,0.80}
\definecolor{DarkSlateGray4}{rgb}{0.32,0.55,0.55}
\definecolor{DarkSlateGray}{rgb}{0.18,0.31,0.31}
\definecolor{DarkSlateGrey}{rgb}{0.18,0.31,0.31}
\definecolor{DarkTurquoise}{rgb}{0.00,0.81,0.82}
\definecolor{DarkViolet}{rgb}{0.58,0.00,0.83}
\definecolor{DeepPink1}{rgb}{1.00,0.08,0.58}
\definecolor{DeepPink2}{rgb}{0.93,0.07,0.54}
\definecolor{DeepPink3}{rgb}{0.80,0.06,0.46}
\definecolor{DeepPink4}{rgb}{0.55,0.04,0.31}
\definecolor{DeepPink}{rgb}{1.00,0.08,0.58}
\definecolor{DeepSkyBlue1}{rgb}{0.00,0.75,1.00}
\definecolor{DeepSkyBlue2}{rgb}{0.00,0.70,0.93}
\definecolor{DeepSkyBlue3}{rgb}{0.00,0.60,0.80}
\definecolor{DeepSkyBlue4}{rgb}{0.00,0.41,0.55}
\definecolor{DeepSkyBlue}{rgb}{0.00,0.75,1.00}
\definecolor{DimGray}{rgb}{0.41,0.41,0.41}
\definecolor{DimGrey}{rgb}{0.41,0.41,0.41}
\definecolor{DodgerBlue1}{rgb}{0.12,0.56,1.00}
\definecolor{DodgerBlue2}{rgb}{0.11,0.53,0.93}
\definecolor{DodgerBlue3}{rgb}{0.09,0.45,0.80}
\definecolor{DodgerBlue4}{rgb}{0.06,0.31,0.55}
\definecolor{DodgerBlue}{rgb}{0.12,0.56,1.00}
\definecolor{FloralWhite}{rgb}{1.00,0.98,0.94}
\definecolor{ForestGreen}{rgb}{0.13,0.55,0.13}
\definecolor{GhostWhite}{rgb}{0.97,0.97,1.00}
\definecolor{GreenYellow}{rgb}{0.68,1.00,0.18}
\definecolor{HotPink1}{rgb}{1.00,0.43,0.71}
\definecolor{HotPink2}{rgb}{0.93,0.42,0.65}
\definecolor{HotPink3}{rgb}{0.80,0.38,0.56}
\definecolor{HotPink4}{rgb}{0.55,0.23,0.38}
\definecolor{HotPink}{rgb}{1.00,0.41,0.71}
\definecolor{IndianRed1}{rgb}{1.00,0.42,0.42}
\definecolor{IndianRed2}{rgb}{0.93,0.39,0.39}
\definecolor{IndianRed3}{rgb}{0.80,0.33,0.33}
\definecolor{IndianRed4}{rgb}{0.55,0.23,0.23}
\definecolor{IndianRed}{rgb}{0.80,0.36,0.36}
\definecolor{LavenderBlush1}{rgb}{1.00,0.94,0.96}
\definecolor{LavenderBlush2}{rgb}{0.93,0.88,0.90}
\definecolor{LavenderBlush3}{rgb}{0.80,0.76,0.77}
\definecolor{LavenderBlush4}{rgb}{0.55,0.51,0.53}
\definecolor{LavenderBlush}{rgb}{1.00,0.94,0.96}
\definecolor{LawnGreen}{rgb}{0.49,0.99,0.00}
\definecolor{LemonChiffon1}{rgb}{1.00,0.98,0.80}
\definecolor{LemonChiffon2}{rgb}{0.93,0.91,0.75}
\definecolor{LemonChiffon3}{rgb}{0.80,0.79,0.65}
\definecolor{LemonChiffon4}{rgb}{0.55,0.54,0.44}
\definecolor{LemonChiffon}{rgb}{1.00,0.98,0.80}
\definecolor{LightBlue1}{rgb}{0.75,0.94,1.00}
\definecolor{LightBlue2}{rgb}{0.70,0.87,0.93}
\definecolor{LightBlue3}{rgb}{0.60,0.75,0.80}
\definecolor{LightBlue4}{rgb}{0.41,0.51,0.55}
\definecolor{LightBlue}{rgb}{0.68,0.85,0.90}
\definecolor{LightCoral}{rgb}{0.94,0.50,0.50}
\definecolor{LightCyan1}{rgb}{0.88,1.00,1.00}
\definecolor{LightCyan2}{rgb}{0.82,0.93,0.93}
\definecolor{LightCyan3}{rgb}{0.71,0.80,0.80}
\definecolor{LightCyan4}{rgb}{0.48,0.55,0.55}
\definecolor{LightCyan}{rgb}{0.88,1.00,1.00}
\definecolor{LightGoldenrod1}{rgb}{1.00,0.93,0.55}
\definecolor{LightGoldenrod2}{rgb}{0.93,0.86,0.51}
\definecolor{LightGoldenrod3}{rgb}{0.80,0.75,0.44}
\definecolor{LightGoldenrod4}{rgb}{0.55,0.51,0.30}
\definecolor{LightGoldenrodYellow}{rgb}{0.98,0.98,0.82}
\definecolor{LightGoldenrod}{rgb}{0.93,0.87,0.51}
\definecolor{LightGray}{rgb}{0.83,0.83,0.83}
\definecolor{LightGreen}{rgb}{0.56,0.93,0.56}
\definecolor{LightGrey}{rgb}{0.83,0.83,0.83}
\definecolor{LightPink1}{rgb}{1.00,0.68,0.73}
\definecolor{LightPink2}{rgb}{0.93,0.64,0.68}
\definecolor{LightPink3}{rgb}{0.80,0.55,0.58}
\definecolor{LightPink4}{rgb}{0.55,0.37,0.40}
\definecolor{LightPink}{rgb}{1.00,0.71,0.76}
\definecolor{LightSalmon1}{rgb}{1.00,0.63,0.48}
\definecolor{LightSalmon2}{rgb}{0.93,0.58,0.45}
\definecolor{LightSalmon3}{rgb}{0.80,0.51,0.38}
\definecolor{LightSalmon4}{rgb}{0.55,0.34,0.26}
\definecolor{LightSalmon}{rgb}{1.00,0.63,0.48}
\definecolor{LightSeaGreen}{rgb}{0.13,0.70,0.67}
\definecolor{LightSkyBlue1}{rgb}{0.69,0.89,1.00}
\definecolor{LightSkyBlue2}{rgb}{0.64,0.83,0.93}
\definecolor{LightSkyBlue3}{rgb}{0.55,0.71,0.80}
\definecolor{LightSkyBlue4}{rgb}{0.38,0.48,0.55}
\definecolor{LightSkyBlue}{rgb}{0.53,0.81,0.98}
\definecolor{LightSlateBlue}{rgb}{0.52,0.44,1.00}
\definecolor{LightSlateGray}{rgb}{0.47,0.53,0.60}
\definecolor{LightSlateGrey}{rgb}{0.47,0.53,0.60}
\definecolor{LightSteelBlue1}{rgb}{0.79,0.88,1.00}
\definecolor{LightSteelBlue2}{rgb}{0.74,0.82,0.93}
\definecolor{LightSteelBlue3}{rgb}{0.64,0.71,0.80}
\definecolor{LightSteelBlue4}{rgb}{0.43,0.48,0.55}
\definecolor{LightSteelBlue}{rgb}{0.69,0.77,0.87}
\definecolor{LightYellow1}{rgb}{1.00,1.00,0.88}
\definecolor{LightYellow2}{rgb}{0.93,0.93,0.82}
\definecolor{LightYellow3}{rgb}{0.80,0.80,0.71}
\definecolor{LightYellow4}{rgb}{0.55,0.55,0.48}
\definecolor{LightYellow}{rgb}{1.00,1.00,0.88}
\definecolor{LimeGreen}{rgb}{0.20,0.80,0.20}
\definecolor{MediumAquamarine}{rgb}{0.40,0.80,0.67}
\definecolor{MediumBlue}{rgb}{0.00,0.00,0.80}
\definecolor{MediumOrchid1}{rgb}{0.88,0.40,1.00}
\definecolor{MediumOrchid2}{rgb}{0.82,0.37,0.93}
\definecolor{MediumOrchid3}{rgb}{0.71,0.32,0.80}
\definecolor{MediumOrchid4}{rgb}{0.48,0.22,0.55}
\definecolor{MediumOrchid}{rgb}{0.73,0.33,0.83}
\definecolor{MediumPurple1}{rgb}{0.67,0.51,1.00}
\definecolor{MediumPurple2}{rgb}{0.62,0.47,0.93}
\definecolor{MediumPurple3}{rgb}{0.54,0.41,0.80}
\definecolor{MediumPurple4}{rgb}{0.36,0.28,0.55}
\definecolor{MediumPurple}{rgb}{0.58,0.44,0.86}
\definecolor{MediumSeaGreen}{rgb}{0.24,0.70,0.44}
\definecolor{MediumSlateBlue}{rgb}{0.48,0.41,0.93}
\definecolor{MediumSpringGreen}{rgb}{0.00,0.98,0.60}
\definecolor{MediumTurquoise}{rgb}{0.28,0.82,0.80}
\definecolor{MediumVioletRed}{rgb}{0.78,0.08,0.52}
\definecolor{MidnightBlue}{rgb}{0.10,0.10,0.44}
\definecolor{MintCream}{rgb}{0.96,1.00,0.98}
\definecolor{MistyRose1}{rgb}{1.00,0.89,0.88}
\definecolor{MistyRose2}{rgb}{0.93,0.84,0.82}
\definecolor{MistyRose3}{rgb}{0.80,0.72,0.71}
\definecolor{MistyRose4}{rgb}{0.55,0.49,0.48}
\definecolor{MistyRose}{rgb}{1.00,0.89,0.88}
\definecolor{NavajoWhite1}{rgb}{1.00,0.87,0.68}
\definecolor{NavajoWhite2}{rgb}{0.93,0.81,0.63}
\definecolor{NavajoWhite3}{rgb}{0.80,0.70,0.55}
\definecolor{NavajoWhite4}{rgb}{0.55,0.47,0.37}
\definecolor{NavajoWhite}{rgb}{1.00,0.87,0.68}
\definecolor{NavyBlue}{rgb}{0.00,0.00,0.50}
\definecolor{OldLace}{rgb}{0.99,0.96,0.90}
\definecolor{OliveDrab1}{rgb}{0.75,1.00,0.24}
\definecolor{OliveDrab2}{rgb}{0.70,0.93,0.23}
\definecolor{OliveDrab3}{rgb}{0.60,0.80,0.20}
\definecolor{OliveDrab4}{rgb}{0.41,0.55,0.13}
\definecolor{OliveDrab}{rgb}{0.42,0.56,0.14}
\definecolor{OrangeRed1}{rgb}{1.00,0.27,0.00}
\definecolor{OrangeRed2}{rgb}{0.93,0.25,0.00}
\definecolor{OrangeRed3}{rgb}{0.80,0.22,0.00}
\definecolor{OrangeRed4}{rgb}{0.55,0.15,0.00}
\definecolor{OrangeRed}{rgb}{1.00,0.27,0.00}
\definecolor{PaleGoldenrod}{rgb}{0.93,0.91,0.67}
\definecolor{PaleGreen1}{rgb}{0.60,1.00,0.60}
\definecolor{PaleGreen2}{rgb}{0.56,0.93,0.56}
\definecolor{PaleGreen3}{rgb}{0.49,0.80,0.49}
\definecolor{PaleGreen4}{rgb}{0.33,0.55,0.33}
\definecolor{PaleGreen}{rgb}{0.60,0.98,0.60}
\definecolor{PaleTurquoise1}{rgb}{0.73,1.00,1.00}
\definecolor{PaleTurquoise2}{rgb}{0.68,0.93,0.93}
\definecolor{PaleTurquoise3}{rgb}{0.59,0.80,0.80}
\definecolor{PaleTurquoise4}{rgb}{0.40,0.55,0.55}
\definecolor{PaleTurquoise}{rgb}{0.69,0.93,0.93}
\definecolor{PaleVioletRed1}{rgb}{1.00,0.51,0.67}
\definecolor{PaleVioletRed2}{rgb}{0.93,0.47,0.62}
\definecolor{PaleVioletRed3}{rgb}{0.80,0.41,0.54}
\definecolor{PaleVioletRed4}{rgb}{0.55,0.28,0.36}
\definecolor{PaleVioletRed}{rgb}{0.86,0.44,0.58}
\definecolor{PapayaWhip}{rgb}{1.00,0.94,0.84}
\definecolor{PeachPuff1}{rgb}{1.00,0.85,0.73}
\definecolor{PeachPuff2}{rgb}{0.93,0.80,0.68}
\definecolor{PeachPuff3}{rgb}{0.80,0.69,0.58}
\definecolor{PeachPuff4}{rgb}{0.55,0.47,0.40}
\definecolor{PeachPuff}{rgb}{1.00,0.85,0.73}
\definecolor{PowderBlue}{rgb}{0.69,0.88,0.90}
\definecolor{RosyBrown1}{rgb}{1.00,0.76,0.76}
\definecolor{RosyBrown2}{rgb}{0.93,0.71,0.71}
\definecolor{RosyBrown3}{rgb}{0.80,0.61,0.61}
\definecolor{RosyBrown4}{rgb}{0.55,0.41,0.41}
\definecolor{RosyBrown}{rgb}{0.74,0.56,0.56}
\definecolor{RoyalBlue1}{rgb}{0.28,0.46,1.00}
\definecolor{RoyalBlue2}{rgb}{0.26,0.43,0.93}
\definecolor{RoyalBlue3}{rgb}{0.23,0.37,0.80}
\definecolor{RoyalBlue4}{rgb}{0.15,0.25,0.55}
\definecolor{RoyalBlue}{rgb}{0.25,0.41,0.88}
\definecolor{SaddleBrown}{rgb}{0.55,0.27,0.07}
\definecolor{SandyBrown}{rgb}{0.96,0.64,0.38}
\definecolor{SeaGreen1}{rgb}{0.33,1.00,0.62}
\definecolor{SeaGreen2}{rgb}{0.31,0.93,0.58}
\definecolor{SeaGreen3}{rgb}{0.26,0.80,0.50}
\definecolor{SeaGreen4}{rgb}{0.18,0.55,0.34}
\definecolor{SeaGreen}{rgb}{0.18,0.55,0.34}
\definecolor{SkyBlue1}{rgb}{0.53,0.81,1.00}
\definecolor{SkyBlue2}{rgb}{0.49,0.75,0.93}
\definecolor{SkyBlue3}{rgb}{0.42,0.65,0.80}
\definecolor{SkyBlue4}{rgb}{0.29,0.44,0.55}
\definecolor{SkyBlue}{rgb}{0.53,0.81,0.92}
\definecolor{SlateBlue1}{rgb}{0.51,0.44,1.00}
\definecolor{SlateBlue2}{rgb}{0.48,0.40,0.93}
\definecolor{SlateBlue3}{rgb}{0.41,0.35,0.80}
\definecolor{SlateBlue4}{rgb}{0.28,0.24,0.55}
\definecolor{SlateBlue}{rgb}{0.42,0.35,0.80}
\definecolor{SlateGray1}{rgb}{0.78,0.89,1.00}
\definecolor{SlateGray2}{rgb}{0.73,0.83,0.93}
\definecolor{SlateGray3}{rgb}{0.62,0.71,0.80}
\definecolor{SlateGray4}{rgb}{0.42,0.48,0.55}
\definecolor{SlateGray}{rgb}{0.44,0.50,0.56}
\definecolor{SlateGrey}{rgb}{0.44,0.50,0.56}
\definecolor{SpringGreen1}{rgb}{0.00,1.00,0.50}
\definecolor{SpringGreen2}{rgb}{0.00,0.93,0.46}
\definecolor{SpringGreen3}{rgb}{0.00,0.80,0.40}
\definecolor{SpringGreen4}{rgb}{0.00,0.55,0.27}
\definecolor{SpringGreen}{rgb}{0.00,1.00,0.50}
\definecolor{SteelBlue1}{rgb}{0.39,0.72,1.00}
\definecolor{SteelBlue2}{rgb}{0.36,0.67,0.93}
\definecolor{SteelBlue3}{rgb}{0.31,0.58,0.80}
\definecolor{SteelBlue4}{rgb}{0.21,0.39,0.55}
\definecolor{SteelBlue}{rgb}{0.27,0.51,0.71}
\definecolor{VioletRed1}{rgb}{1.00,0.24,0.59}
\definecolor{VioletRed2}{rgb}{0.93,0.23,0.55}
\definecolor{VioletRed3}{rgb}{0.80,0.20,0.47}
\definecolor{VioletRed4}{rgb}{0.55,0.13,0.32}
\definecolor{VioletRed}{rgb}{0.82,0.13,0.56}
\definecolor{WhiteSmoke}{rgb}{0.96,0.96,0.96}
\definecolor{YellowGreen}{rgb}{0.60,0.80,0.20}
\definecolor{aliceblue}{rgb}{0.94,0.97,1.00}
\definecolor{antiquewhite}{rgb}{0.98,0.92,0.84}
\definecolor{aquamarine1}{rgb}{0.50,1.00,0.83}
\definecolor{aquamarine2}{rgb}{0.46,0.93,0.78}
\definecolor{aquamarine3}{rgb}{0.40,0.80,0.67}
\definecolor{aquamarine4}{rgb}{0.27,0.55,0.45}
\definecolor{aquamarine}{rgb}{0.50,1.00,0.83}
\definecolor{azure1}{rgb}{0.94,1.00,1.00}
\definecolor{azure2}{rgb}{0.88,0.93,0.93}
\definecolor{azure3}{rgb}{0.76,0.80,0.80}
\definecolor{azure4}{rgb}{0.51,0.55,0.55}
\definecolor{azure}{rgb}{0.94,1.00,1.00}
\definecolor{beige}{rgb}{0.96,0.96,0.86}
\definecolor{bisque1}{rgb}{1.00,0.89,0.77}
\definecolor{bisque2}{rgb}{0.93,0.84,0.72}
\definecolor{bisque3}{rgb}{0.80,0.72,0.62}
\definecolor{bisque4}{rgb}{0.55,0.49,0.42}
\definecolor{bisque}{rgb}{1.00,0.89,0.77}
\definecolor{black}{rgb}{0.00,0.00,0.00}
\definecolor{blanchedalmond}{rgb}{1.00,0.92,0.80}
\definecolor{blue1}{rgb}{0.00,0.00,1.00}
\definecolor{blue2}{rgb}{0.00,0.00,0.93}
\definecolor{blue3}{rgb}{0.00,0.00,0.80}
\definecolor{blue4}{rgb}{0.00,0.00,0.55}
\definecolor{blueviolet}{rgb}{0.54,0.17,0.89}
\definecolor{blue}{rgb}{0.00,0.00,1.00}
\definecolor{brown1}{rgb}{1.00,0.25,0.25}
\definecolor{brown2}{rgb}{0.93,0.23,0.23}
\definecolor{brown3}{rgb}{0.80,0.20,0.20}
\definecolor{brown4}{rgb}{0.55,0.14,0.14}
\definecolor{brown}{rgb}{0.65,0.16,0.16}
\definecolor{burlywood1}{rgb}{1.00,0.83,0.61}
\definecolor{burlywood2}{rgb}{0.93,0.77,0.57}
\definecolor{burlywood3}{rgb}{0.80,0.67,0.49}
\definecolor{burlywood4}{rgb}{0.55,0.45,0.33}
\definecolor{burlywood}{rgb}{0.87,0.72,0.53}
\definecolor{cadetblue}{rgb}{0.37,0.62,0.63}
\definecolor{chartreuse1}{rgb}{0.50,1.00,0.00}
\definecolor{chartreuse2}{rgb}{0.46,0.93,0.00}
\definecolor{chartreuse3}{rgb}{0.40,0.80,0.00}
\definecolor{chartreuse4}{rgb}{0.27,0.55,0.00}
\definecolor{chartreuse}{rgb}{0.50,1.00,0.00}
\definecolor{chocolate1}{rgb}{1.00,0.50,0.14}
\definecolor{chocolate2}{rgb}{0.93,0.46,0.13}
\definecolor{chocolate3}{rgb}{0.80,0.40,0.11}
\definecolor{chocolate4}{rgb}{0.55,0.27,0.07}
\definecolor{chocolate}{rgb}{0.82,0.41,0.12}
\definecolor{coral1}{rgb}{1.00,0.45,0.34}
\definecolor{coral2}{rgb}{0.93,0.42,0.31}
\definecolor{coral3}{rgb}{0.80,0.36,0.27}
\definecolor{coral4}{rgb}{0.55,0.24,0.18}
\definecolor{coral}{rgb}{1.00,0.50,0.31}
\definecolor{cornflowerblue}{rgb}{0.39,0.58,0.93}
\definecolor{cornsilk1}{rgb}{1.00,0.97,0.86}
\definecolor{cornsilk2}{rgb}{0.93,0.91,0.80}
\definecolor{cornsilk3}{rgb}{0.80,0.78,0.69}
\definecolor{cornsilk4}{rgb}{0.55,0.53,0.47}
\definecolor{cornsilk}{rgb}{1.00,0.97,0.86}
\definecolor{cyan1}{rgb}{0.00,1.00,1.00}
\definecolor{cyan2}{rgb}{0.00,0.93,0.93}
\definecolor{cyan3}{rgb}{0.00,0.80,0.80}
\definecolor{cyan4}{rgb}{0.00,0.55,0.55}
\definecolor{cyan}{rgb}{0.00,1.00,1.00}
\definecolor{darkblue}{rgb}{0.00,0.00,0.55}
\definecolor{darkcyan}{rgb}{0.00,0.55,0.55}
\definecolor{darkgoldenrod}{rgb}{0.72,0.53,0.04}
\definecolor{darkgray}{rgb}{0.66,0.66,0.66}
\definecolor{darkgreen}{rgb}{0.00,0.39,0.00}
\definecolor{darkgrey}{rgb}{0.66,0.66,0.66}
\definecolor{darkkhaki}{rgb}{0.74,0.72,0.42}
\definecolor{darkmagenta}{rgb}{0.55,0.00,0.55}
\definecolor{darkolive}{rgb}{0.33,0.42,0.18}
\definecolor{darkorange}{rgb}{1.00,0.55,0.00}
\definecolor{darkorchid}{rgb}{0.60,0.20,0.80}
\definecolor{darkred}{rgb}{0.55,0.00,0.00}
\definecolor{darksalmon}{rgb}{0.91,0.59,0.48}
\definecolor{darksea}{rgb}{0.56,0.74,0.56}
\definecolor{darkslate}{rgb}{0.18,0.31,0.31}
\definecolor{darkslate}{rgb}{0.18,0.31,0.31}
\definecolor{darkslate}{rgb}{0.28,0.24,0.55}
\definecolor{darkturquoise}{rgb}{0.00,0.81,0.82}
\definecolor{darkviolet}{rgb}{0.58,0.00,0.83}
\definecolor{deeppink}{rgb}{1.00,0.08,0.58}
\definecolor{deepsky}{rgb}{0.00,0.75,1.00}
\definecolor{dimgray}{rgb}{0.41,0.41,0.41}
\definecolor{dimgrey}{rgb}{0.41,0.41,0.41}
\definecolor{dodgerblue}{rgb}{0.12,0.56,1.00}
\definecolor{firebrick1}{rgb}{1.00,0.19,0.19}
\definecolor{firebrick2}{rgb}{0.93,0.17,0.17}
\definecolor{firebrick3}{rgb}{0.80,0.15,0.15}
\definecolor{firebrick4}{rgb}{0.55,0.10,0.10}
\definecolor{firebrick}{rgb}{0.70,0.13,0.13}
\definecolor{floralwhite}{rgb}{1.00,0.98,0.94}
\definecolor{forestgreen}{rgb}{0.13,0.55,0.13}
\definecolor{gainsboro}{rgb}{0.86,0.86,0.86}
\definecolor{ghostwhite}{rgb}{0.97,0.97,1.00}
\definecolor{gold1}{rgb}{1.00,0.84,0.00}
\definecolor{gold2}{rgb}{0.93,0.79,0.00}
\definecolor{gold3}{rgb}{0.80,0.68,0.00}
\definecolor{gold4}{rgb}{0.55,0.46,0.00}
\definecolor{goldenrod1}{rgb}{1.00,0.76,0.15}
\definecolor{goldenrod2}{rgb}{0.93,0.71,0.13}
\definecolor{goldenrod3}{rgb}{0.80,0.61,0.11}
\definecolor{goldenrod4}{rgb}{0.55,0.41,0.08}
\definecolor{goldenrod}{rgb}{0.85,0.65,0.13}
\definecolor{gold}{rgb}{1.00,0.84,0.00}
\definecolor{gray0}{rgb}{0.00,0.00,0.00}
\definecolor{gray100}{rgb}{1.00,1.00,1.00}
\definecolor{gray10}{rgb}{0.10,0.10,0.10}
\definecolor{gray11}{rgb}{0.11,0.11,0.11}
\definecolor{gray12}{rgb}{0.12,0.12,0.12}
\definecolor{gray13}{rgb}{0.13,0.13,0.13}
\definecolor{gray14}{rgb}{0.14,0.14,0.14}
\definecolor{gray15}{rgb}{0.15,0.15,0.15}
\definecolor{gray16}{rgb}{0.16,0.16,0.16}
\definecolor{gray17}{rgb}{0.17,0.17,0.17}
\definecolor{gray18}{rgb}{0.18,0.18,0.18}
\definecolor{gray19}{rgb}{0.19,0.19,0.19}
\definecolor{gray1}{rgb}{0.01,0.01,0.01}
\definecolor{gray20}{rgb}{0.20,0.20,0.20}
\definecolor{gray21}{rgb}{0.21,0.21,0.21}
\definecolor{gray22}{rgb}{0.22,0.22,0.22}
\definecolor{gray23}{rgb}{0.23,0.23,0.23}
\definecolor{gray24}{rgb}{0.24,0.24,0.24}
\definecolor{gray25}{rgb}{0.25,0.25,0.25}
\definecolor{gray26}{rgb}{0.26,0.26,0.26}
\definecolor{gray27}{rgb}{0.27,0.27,0.27}
\definecolor{gray28}{rgb}{0.28,0.28,0.28}
\definecolor{gray29}{rgb}{0.29,0.29,0.29}
\definecolor{gray2}{rgb}{0.02,0.02,0.02}
\definecolor{gray30}{rgb}{0.30,0.30,0.30}
\definecolor{gray31}{rgb}{0.31,0.31,0.31}
\definecolor{gray32}{rgb}{0.32,0.32,0.32}
\definecolor{gray33}{rgb}{0.33,0.33,0.33}
\definecolor{gray34}{rgb}{0.34,0.34,0.34}
\definecolor{gray35}{rgb}{0.35,0.35,0.35}
\definecolor{gray36}{rgb}{0.36,0.36,0.36}
\definecolor{gray37}{rgb}{0.37,0.37,0.37}
\definecolor{gray38}{rgb}{0.38,0.38,0.38}
\definecolor{gray39}{rgb}{0.39,0.39,0.39}
\definecolor{gray3}{rgb}{0.03,0.03,0.03}
\definecolor{gray40}{rgb}{0.40,0.40,0.40}
\definecolor{gray41}{rgb}{0.41,0.41,0.41}
\definecolor{gray42}{rgb}{0.42,0.42,0.42}
\definecolor{gray43}{rgb}{0.43,0.43,0.43}
\definecolor{gray44}{rgb}{0.44,0.44,0.44}
\definecolor{gray45}{rgb}{0.45,0.45,0.45}
\definecolor{gray46}{rgb}{0.46,0.46,0.46}
\definecolor{gray47}{rgb}{0.47,0.47,0.47}
\definecolor{gray48}{rgb}{0.48,0.48,0.48}
\definecolor{gray49}{rgb}{0.49,0.49,0.49}
\definecolor{gray4}{rgb}{0.04,0.04,0.04}
\definecolor{gray50}{rgb}{0.50,0.50,0.50}
\definecolor{gray51}{rgb}{0.51,0.51,0.51}
\definecolor{gray52}{rgb}{0.52,0.52,0.52}
\definecolor{gray53}{rgb}{0.53,0.53,0.53}
\definecolor{gray54}{rgb}{0.54,0.54,0.54}
\definecolor{gray55}{rgb}{0.55,0.55,0.55}
\definecolor{gray56}{rgb}{0.56,0.56,0.56}
\definecolor{gray57}{rgb}{0.57,0.57,0.57}
\definecolor{gray58}{rgb}{0.58,0.58,0.58}
\definecolor{gray59}{rgb}{0.59,0.59,0.59}
\definecolor{gray5}{rgb}{0.05,0.05,0.05}
\definecolor{gray60}{rgb}{0.60,0.60,0.60}
\definecolor{gray61}{rgb}{0.61,0.61,0.61}
\definecolor{gray62}{rgb}{0.62,0.62,0.62}
\definecolor{gray63}{rgb}{0.63,0.63,0.63}
\definecolor{gray64}{rgb}{0.64,0.64,0.64}
\definecolor{gray65}{rgb}{0.65,0.65,0.65}
\definecolor{gray66}{rgb}{0.66,0.66,0.66}
\definecolor{gray67}{rgb}{0.67,0.67,0.67}
\definecolor{gray68}{rgb}{0.68,0.68,0.68}
\definecolor{gray69}{rgb}{0.69,0.69,0.69}
\definecolor{gray6}{rgb}{0.06,0.06,0.06}
\definecolor{gray70}{rgb}{0.70,0.70,0.70}
\definecolor{gray71}{rgb}{0.71,0.71,0.71}
\definecolor{gray72}{rgb}{0.72,0.72,0.72}
\definecolor{gray73}{rgb}{0.73,0.73,0.73}
\definecolor{gray74}{rgb}{0.74,0.74,0.74}
\definecolor{gray75}{rgb}{0.75,0.75,0.75}
\definecolor{gray76}{rgb}{0.76,0.76,0.76}
\definecolor{gray77}{rgb}{0.77,0.77,0.77}
\definecolor{gray78}{rgb}{0.78,0.78,0.78}
\definecolor{gray79}{rgb}{0.79,0.79,0.79}
\definecolor{gray7}{rgb}{0.07,0.07,0.07}
\definecolor{gray80}{rgb}{0.80,0.80,0.80}
\definecolor{gray81}{rgb}{0.81,0.81,0.81}
\definecolor{gray82}{rgb}{0.82,0.82,0.82}
\definecolor{gray83}{rgb}{0.83,0.83,0.83}
\definecolor{gray84}{rgb}{0.84,0.84,0.84}
\definecolor{gray85}{rgb}{0.85,0.85,0.85}
\definecolor{gray86}{rgb}{0.86,0.86,0.86}
\definecolor{gray87}{rgb}{0.87,0.87,0.87}
\definecolor{gray88}{rgb}{0.88,0.88,0.88}
\definecolor{gray89}{rgb}{0.89,0.89,0.89}
\definecolor{gray8}{rgb}{0.08,0.08,0.08}
\definecolor{gray90}{rgb}{0.90,0.90,0.90}
\definecolor{gray91}{rgb}{0.91,0.91,0.91}
\definecolor{gray92}{rgb}{0.92,0.92,0.92}
\definecolor{gray93}{rgb}{0.93,0.93,0.93}
\definecolor{gray94}{rgb}{0.94,0.94,0.94}
\definecolor{gray95}{rgb}{0.95,0.95,0.95}
\definecolor{gray96}{rgb}{0.96,0.96,0.96}
\definecolor{gray97}{rgb}{0.97,0.97,0.97}
\definecolor{gray98}{rgb}{0.98,0.98,0.98}
\definecolor{gray99}{rgb}{0.99,0.99,0.99}
\definecolor{gray9}{rgb}{0.09,0.09,0.09}
\definecolor{gray}{rgb}{0.75,0.75,0.75}
\definecolor{green1}{rgb}{0.00,1.00,0.00}
\definecolor{green2}{rgb}{0.00,0.93,0.00}
\definecolor{green3}{rgb}{0.00,0.80,0.00}
\definecolor{green4}{rgb}{0.00,0.55,0.00}
\definecolor{greenyellow}{rgb}{0.68,1.00,0.18}
\definecolor{green}{rgb}{0.00,1.00,0.00}
\definecolor{grey0}{rgb}{0.00,0.00,0.00}
\definecolor{grey100}{rgb}{1.00,1.00,1.00}
\definecolor{grey10}{rgb}{0.10,0.10,0.10}
\definecolor{grey11}{rgb}{0.11,0.11,0.11}
\definecolor{grey12}{rgb}{0.12,0.12,0.12}
\definecolor{grey13}{rgb}{0.13,0.13,0.13}
\definecolor{grey14}{rgb}{0.14,0.14,0.14}
\definecolor{grey15}{rgb}{0.15,0.15,0.15}
\definecolor{grey16}{rgb}{0.16,0.16,0.16}
\definecolor{grey17}{rgb}{0.17,0.17,0.17}
\definecolor{grey18}{rgb}{0.18,0.18,0.18}
\definecolor{grey19}{rgb}{0.19,0.19,0.19}
\definecolor{grey1}{rgb}{0.01,0.01,0.01}
\definecolor{grey20}{rgb}{0.20,0.20,0.20}
\definecolor{grey21}{rgb}{0.21,0.21,0.21}
\definecolor{grey22}{rgb}{0.22,0.22,0.22}
\definecolor{grey23}{rgb}{0.23,0.23,0.23}
\definecolor{grey24}{rgb}{0.24,0.24,0.24}
\definecolor{grey25}{rgb}{0.25,0.25,0.25}
\definecolor{grey26}{rgb}{0.26,0.26,0.26}
\definecolor{grey27}{rgb}{0.27,0.27,0.27}
\definecolor{grey28}{rgb}{0.28,0.28,0.28}
\definecolor{grey29}{rgb}{0.29,0.29,0.29}
\definecolor{grey2}{rgb}{0.02,0.02,0.02}
\definecolor{grey30}{rgb}{0.30,0.30,0.30}
\definecolor{grey31}{rgb}{0.31,0.31,0.31}
\definecolor{grey32}{rgb}{0.32,0.32,0.32}
\definecolor{grey33}{rgb}{0.33,0.33,0.33}
\definecolor{grey34}{rgb}{0.34,0.34,0.34}
\definecolor{grey35}{rgb}{0.35,0.35,0.35}
\definecolor{grey36}{rgb}{0.36,0.36,0.36}
\definecolor{grey37}{rgb}{0.37,0.37,0.37}
\definecolor{grey38}{rgb}{0.38,0.38,0.38}
\definecolor{grey39}{rgb}{0.39,0.39,0.39}
\definecolor{grey3}{rgb}{0.03,0.03,0.03}
\definecolor{grey40}{rgb}{0.40,0.40,0.40}
\definecolor{grey41}{rgb}{0.41,0.41,0.41}
\definecolor{grey42}{rgb}{0.42,0.42,0.42}
\definecolor{grey43}{rgb}{0.43,0.43,0.43}
\definecolor{grey44}{rgb}{0.44,0.44,0.44}
\definecolor{grey45}{rgb}{0.45,0.45,0.45}
\definecolor{grey46}{rgb}{0.46,0.46,0.46}
\definecolor{grey47}{rgb}{0.47,0.47,0.47}
\definecolor{grey48}{rgb}{0.48,0.48,0.48}
\definecolor{grey49}{rgb}{0.49,0.49,0.49}
\definecolor{grey4}{rgb}{0.04,0.04,0.04}
\definecolor{grey50}{rgb}{0.50,0.50,0.50}
\definecolor{grey51}{rgb}{0.51,0.51,0.51}
\definecolor{grey52}{rgb}{0.52,0.52,0.52}
\definecolor{grey53}{rgb}{0.53,0.53,0.53}
\definecolor{grey54}{rgb}{0.54,0.54,0.54}
\definecolor{grey55}{rgb}{0.55,0.55,0.55}
\definecolor{grey56}{rgb}{0.56,0.56,0.56}
\definecolor{grey57}{rgb}{0.57,0.57,0.57}
\definecolor{grey58}{rgb}{0.58,0.58,0.58}
\definecolor{grey59}{rgb}{0.59,0.59,0.59}
\definecolor{grey5}{rgb}{0.05,0.05,0.05}
\definecolor{grey60}{rgb}{0.60,0.60,0.60}
\definecolor{grey61}{rgb}{0.61,0.61,0.61}
\definecolor{grey62}{rgb}{0.62,0.62,0.62}
\definecolor{grey63}{rgb}{0.63,0.63,0.63}
\definecolor{grey64}{rgb}{0.64,0.64,0.64}
\definecolor{grey65}{rgb}{0.65,0.65,0.65}
\definecolor{grey66}{rgb}{0.66,0.66,0.66}
\definecolor{grey67}{rgb}{0.67,0.67,0.67}
\definecolor{grey68}{rgb}{0.68,0.68,0.68}
\definecolor{grey69}{rgb}{0.69,0.69,0.69}
\definecolor{grey6}{rgb}{0.06,0.06,0.06}
\definecolor{grey70}{rgb}{0.70,0.70,0.70}
\definecolor{grey71}{rgb}{0.71,0.71,0.71}
\definecolor{grey72}{rgb}{0.72,0.72,0.72}
\definecolor{grey73}{rgb}{0.73,0.73,0.73}
\definecolor{grey74}{rgb}{0.74,0.74,0.74}
\definecolor{grey75}{rgb}{0.75,0.75,0.75}
\definecolor{grey76}{rgb}{0.76,0.76,0.76}
\definecolor{grey77}{rgb}{0.77,0.77,0.77}
\definecolor{grey78}{rgb}{0.78,0.78,0.78}
\definecolor{grey79}{rgb}{0.79,0.79,0.79}
\definecolor{grey7}{rgb}{0.07,0.07,0.07}
\definecolor{grey80}{rgb}{0.80,0.80,0.80}
\definecolor{grey81}{rgb}{0.81,0.81,0.81}
\definecolor{grey82}{rgb}{0.82,0.82,0.82}
\definecolor{grey83}{rgb}{0.83,0.83,0.83}
\definecolor{grey84}{rgb}{0.84,0.84,0.84}
\definecolor{grey85}{rgb}{0.85,0.85,0.85}
\definecolor{grey86}{rgb}{0.86,0.86,0.86}
\definecolor{grey87}{rgb}{0.87,0.87,0.87}
\definecolor{grey88}{rgb}{0.88,0.88,0.88}
\definecolor{grey89}{rgb}{0.89,0.89,0.89}
\definecolor{grey8}{rgb}{0.08,0.08,0.08}
\definecolor{grey90}{rgb}{0.90,0.90,0.90}
\definecolor{grey91}{rgb}{0.91,0.91,0.91}
\definecolor{grey92}{rgb}{0.92,0.92,0.92}
\definecolor{grey93}{rgb}{0.93,0.93,0.93}
\definecolor{grey94}{rgb}{0.94,0.94,0.94}
\definecolor{grey95}{rgb}{0.95,0.95,0.95}
\definecolor{grey96}{rgb}{0.96,0.96,0.96}
\definecolor{grey97}{rgb}{0.97,0.97,0.97}
\definecolor{grey98}{rgb}{0.98,0.98,0.98}
\definecolor{grey99}{rgb}{0.99,0.99,0.99}
\definecolor{grey9}{rgb}{0.09,0.09,0.09}
\definecolor{grey}{rgb}{0.75,0.75,0.75}
\definecolor{honeydew1}{rgb}{0.94,1.00,0.94}
\definecolor{honeydew2}{rgb}{0.88,0.93,0.88}
\definecolor{honeydew3}{rgb}{0.76,0.80,0.76}
\definecolor{honeydew4}{rgb}{0.51,0.55,0.51}
\definecolor{honeydew}{rgb}{0.94,1.00,0.94}
\definecolor{hotpink}{rgb}{1.00,0.41,0.71}
\definecolor{indianred}{rgb}{0.80,0.36,0.36}
\definecolor{ivory1}{rgb}{1.00,1.00,0.94}
\definecolor{ivory2}{rgb}{0.93,0.93,0.88}
\definecolor{ivory3}{rgb}{0.80,0.80,0.76}
\definecolor{ivory4}{rgb}{0.55,0.55,0.51}
\definecolor{ivory}{rgb}{1.00,1.00,0.94}
\definecolor{khaki1}{rgb}{1.00,0.96,0.56}
\definecolor{khaki2}{rgb}{0.93,0.90,0.52}
\definecolor{khaki3}{rgb}{0.80,0.78,0.45}
\definecolor{khaki4}{rgb}{0.55,0.53,0.31}
\definecolor{khaki}{rgb}{0.94,0.90,0.55}
\definecolor{lavenderblush}{rgb}{1.00,0.94,0.96}
\definecolor{lavender}{rgb}{0.90,0.90,0.98}
\definecolor{lawngreen}{rgb}{0.49,0.99,0.00}
\definecolor{lemonchiffon}{rgb}{1.00,0.98,0.80}
\definecolor{lightblue}{rgb}{0.68,0.85,0.90}
\definecolor{lightcoral}{rgb}{0.94,0.50,0.50}
\definecolor{lightcyan}{rgb}{0.88,1.00,1.00}
\definecolor{lightgoldenrod}{rgb}{0.93,0.87,0.51}
\definecolor{lightgoldenrod}{rgb}{0.98,0.98,0.82}
\definecolor{lightgray}{rgb}{0.83,0.83,0.83}
\definecolor{lightgreen}{rgb}{0.56,0.93,0.56}
\definecolor{lightgrey}{rgb}{0.83,0.83,0.83}
\definecolor{lightpink}{rgb}{1.00,0.71,0.76}
\definecolor{lightsalmon}{rgb}{1.00,0.63,0.48}
\definecolor{lightsea}{rgb}{0.13,0.70,0.67}
\definecolor{lightsky}{rgb}{0.53,0.81,0.98}
\definecolor{lightslate}{rgb}{0.47,0.53,0.60}
\definecolor{lightslate}{rgb}{0.47,0.53,0.60}
\definecolor{lightslate}{rgb}{0.52,0.44,1.00}
\definecolor{lightsteel}{rgb}{0.69,0.77,0.87}
\definecolor{lightyellow}{rgb}{1.00,1.00,0.88}
\definecolor{limegreen}{rgb}{0.20,0.80,0.20}
\definecolor{linen}{rgb}{0.98,0.94,0.90}
\definecolor{magenta1}{rgb}{1.00,0.00,1.00}
\definecolor{magenta2}{rgb}{0.93,0.00,0.93}
\definecolor{magenta3}{rgb}{0.80,0.00,0.80}
\definecolor{magenta4}{rgb}{0.55,0.00,0.55}
\definecolor{magenta}{rgb}{1.00,0.00,1.00}
\definecolor{maroon1}{rgb}{1.00,0.20,0.70}
\definecolor{maroon2}{rgb}{0.93,0.19,0.65}
\definecolor{maroon3}{rgb}{0.80,0.16,0.56}
\definecolor{maroon4}{rgb}{0.55,0.11,0.38}
\definecolor{maroon}{rgb}{0.69,0.19,0.38}
\definecolor{mediumaquamarine}{rgb}{0.40,0.80,0.67}
\definecolor{mediumblue}{rgb}{0.00,0.00,0.80}
\definecolor{mediumorchid}{rgb}{0.73,0.33,0.83}
\definecolor{mediumpurple}{rgb}{0.58,0.44,0.86}
\definecolor{mediumsea}{rgb}{0.24,0.70,0.44}
\definecolor{mediumslate}{rgb}{0.48,0.41,0.93}
\definecolor{mediumspring}{rgb}{0.00,0.98,0.60}
\definecolor{mediumturquoise}{rgb}{0.28,0.82,0.80}
\definecolor{mediumviolet}{rgb}{0.78,0.08,0.52}
\definecolor{midnightblue}{rgb}{0.10,0.10,0.44}
\definecolor{mintcream}{rgb}{0.96,1.00,0.98}
\definecolor{mistyrose}{rgb}{1.00,0.89,0.88}
\definecolor{moccasin}{rgb}{1.00,0.89,0.71}
\definecolor{navajowhite}{rgb}{1.00,0.87,0.68}
\definecolor{navyblue}{rgb}{0.00,0.00,0.50}
\definecolor{navy}{rgb}{0.00,0.00,0.50}
\definecolor{oldlace}{rgb}{0.99,0.96,0.90}
\definecolor{olivedrab}{rgb}{0.42,0.56,0.14}
\definecolor{orange1}{rgb}{1.00,0.65,0.00}
\definecolor{orange2}{rgb}{0.93,0.60,0.00}
\definecolor{orange3}{rgb}{0.80,0.52,0.00}
\definecolor{orange4}{rgb}{0.55,0.35,0.00}
\definecolor{orangered}{rgb}{1.00,0.27,0.00}
\definecolor{orange}{rgb}{1.00,0.65,0.00}
\definecolor{orchid1}{rgb}{1.00,0.51,0.98}
\definecolor{orchid2}{rgb}{0.93,0.48,0.91}
\definecolor{orchid3}{rgb}{0.80,0.41,0.79}
\definecolor{orchid4}{rgb}{0.55,0.28,0.54}
\definecolor{orchid}{rgb}{0.85,0.44,0.84}
\definecolor{palegoldenrod}{rgb}{0.93,0.91,0.67}
\definecolor{palegreen}{rgb}{0.60,0.98,0.60}
\definecolor{paleturquoise}{rgb}{0.69,0.93,0.93}
\definecolor{paleviolet}{rgb}{0.86,0.44,0.58}
\definecolor{papayawhip}{rgb}{1.00,0.94,0.84}
\definecolor{peachpuff}{rgb}{1.00,0.85,0.73}
\definecolor{peru}{rgb}{0.80,0.52,0.25}
\definecolor{pink1}{rgb}{1.00,0.71,0.77}
\definecolor{pink2}{rgb}{0.93,0.66,0.72}
\definecolor{pink3}{rgb}{0.80,0.57,0.62}
\definecolor{pink4}{rgb}{0.55,0.39,0.42}
\definecolor{pink}{rgb}{1.00,0.75,0.80}
\definecolor{plum1}{rgb}{1.00,0.73,1.00}
\definecolor{plum2}{rgb}{0.93,0.68,0.93}
\definecolor{plum3}{rgb}{0.80,0.59,0.80}
\definecolor{plum4}{rgb}{0.55,0.40,0.55}
\definecolor{plum}{rgb}{0.87,0.63,0.87}
\definecolor{powderblue}{rgb}{0.69,0.88,0.90}
\definecolor{purple1}{rgb}{0.61,0.19,1.00}
\definecolor{purple2}{rgb}{0.57,0.17,0.93}
\definecolor{purple3}{rgb}{0.49,0.15,0.80}
\definecolor{purple4}{rgb}{0.33,0.10,0.55}
\definecolor{purple}{rgb}{0.63,0.13,0.94}
\definecolor{red1}{rgb}{1.00,0.00,0.00}
\definecolor{red2}{rgb}{0.93,0.00,0.00}
\definecolor{red3}{rgb}{0.80,0.00,0.00}
\definecolor{red4}{rgb}{0.55,0.00,0.00}
\definecolor{red}{rgb}{1.00,0.00,0.00}
\definecolor{rosybrown}{rgb}{0.74,0.56,0.56}
\definecolor{royalblue}{rgb}{0.25,0.41,0.88}
\definecolor{saddlebrown}{rgb}{0.55,0.27,0.07}
\definecolor{salmon1}{rgb}{1.00,0.55,0.41}
\definecolor{salmon2}{rgb}{0.93,0.51,0.38}
\definecolor{salmon3}{rgb}{0.80,0.44,0.33}
\definecolor{salmon4}{rgb}{0.55,0.30,0.22}
\definecolor{salmon}{rgb}{0.98,0.50,0.45}
\definecolor{sandybrown}{rgb}{0.96,0.64,0.38}
\definecolor{seagreen}{rgb}{0.18,0.55,0.34}
\definecolor{seashell1}{rgb}{1.00,0.96,0.93}
\definecolor{seashell2}{rgb}{0.93,0.90,0.87}
\definecolor{seashell3}{rgb}{0.80,0.77,0.75}
\definecolor{seashell4}{rgb}{0.55,0.53,0.51}
\definecolor{seashell}{rgb}{1.00,0.96,0.93}
\definecolor{sienna1}{rgb}{1.00,0.51,0.28}
\definecolor{sienna2}{rgb}{0.93,0.47,0.26}
\definecolor{sienna3}{rgb}{0.80,0.41,0.22}
\definecolor{sienna4}{rgb}{0.55,0.28,0.15}
\definecolor{sienna}{rgb}{0.63,0.32,0.18}
\definecolor{skyblue}{rgb}{0.53,0.81,0.92}
\definecolor{slateblue}{rgb}{0.42,0.35,0.80}
\definecolor{slategray}{rgb}{0.44,0.50,0.56}
\definecolor{slategrey}{rgb}{0.44,0.50,0.56}
\definecolor{snow1}{rgb}{1.00,0.98,0.98}
\definecolor{snow2}{rgb}{0.93,0.91,0.91}
\definecolor{snow3}{rgb}{0.80,0.79,0.79}
\definecolor{snow4}{rgb}{0.55,0.54,0.54}
\definecolor{snow}{rgb}{1.00,0.98,0.98}
\definecolor{springgreen}{rgb}{0.00,1.00,0.50}
\definecolor{steelblue}{rgb}{0.27,0.51,0.71}
\definecolor{tan1}{rgb}{1.00,0.65,0.31}
\definecolor{tan2}{rgb}{0.93,0.60,0.29}
\definecolor{tan3}{rgb}{0.80,0.52,0.25}
\definecolor{tan4}{rgb}{0.55,0.35,0.17}
\definecolor{tan}{rgb}{0.82,0.71,0.55}
\definecolor{thistle1}{rgb}{1.00,0.88,1.00}
\definecolor{thistle2}{rgb}{0.93,0.82,0.93}
\definecolor{thistle3}{rgb}{0.80,0.71,0.80}
\definecolor{thistle4}{rgb}{0.55,0.48,0.55}
\definecolor{thistle}{rgb}{0.85,0.75,0.85}
\definecolor{tomato1}{rgb}{1.00,0.39,0.28}
\definecolor{tomato2}{rgb}{0.93,0.36,0.26}
\definecolor{tomato3}{rgb}{0.80,0.31,0.22}
\definecolor{tomato4}{rgb}{0.55,0.21,0.15}
\definecolor{tomato}{rgb}{1.00,0.39,0.28}
\definecolor{turquoise1}{rgb}{0.00,0.96,1.00}
\definecolor{turquoise2}{rgb}{0.00,0.90,0.93}
\definecolor{turquoise3}{rgb}{0.00,0.77,0.80}
\definecolor{turquoise4}{rgb}{0.00,0.53,0.55}
\definecolor{turquoise}{rgb}{0.25,0.88,0.82}
\definecolor{violetred}{rgb}{0.82,0.13,0.56}
\definecolor{violet}{rgb}{0.93,0.51,0.93}
\definecolor{wheat1}{rgb}{1.00,0.91,0.73}
\definecolor{wheat2}{rgb}{0.93,0.85,0.68}
\definecolor{wheat3}{rgb}{0.80,0.73,0.59}
\definecolor{wheat4}{rgb}{0.55,0.49,0.40}
\definecolor{wheat}{rgb}{0.96,0.87,0.70}
\definecolor{whitesmoke}{rgb}{0.96,0.96,0.96}
\definecolor{white}{rgb}{1.00,1.00,1.00}
\definecolor{yellow1}{rgb}{1.00,1.00,0.00}
\definecolor{yellow2}{rgb}{0.93,0.93,0.00}
\definecolor{yellow3}{rgb}{0.80,0.80,0.00}
\definecolor{yellow4}{rgb}{0.55,0.55,0.00}
\definecolor{yellowgreen}{rgb}{0.60,0.80,0.20}
\definecolor{yellow}{rgb}{1.00,1.00,0.00}

\usepackage{amsmath}
\usepackage{amssymb}
\usepackage[dvips]{graphicx}
\usepackage{amsmath,amsfonts,amssymb,aas_macros}

\usepackage[normalem]{ulem} 
\usepackage{epstopdf}
\usepackage{epsfig}

\newcommand{\consistenttree}{\textsc{Consistent Trees}}

\newcommand{\mergertree}{\textsc{MergerTree}}

\newcommand{\jmerger}{\textsc{JMerge}}

\newcommand{\sublink}{\textsc{SubLink}}
\newcommand{\treemaker}{\textsc{TreeMaker}}
\newcommand{\velociraptor}{\textsc{VELOCIraptor}}

\newcommand{\ahf}{\textsc{AHF}}

\newcommand{\subfind}{\textsc{SUBFIND}}

\newcommand{\rockstar}{\textsc{Rockstar}}
\newcommand{\hbt}{\textsc{HBT}}
\newcommand{\hbttree}{\textsc{HBTtree}}
\newcommand{\hbthalo}{\textsc{HBThalo}}

\newcommand{\stf}{\textsc{STF}}

\newcommand{\Eq}[1]{Equation~\ref{#1}}
\newcommand{\Fig}[1]{Figure~\ref{#1}}

\newcommand{\Sec}[1]{Section~\ref{#1}}
\newcommand{\Tab}[1]{Table~\ref{#1}}
\newcommand{\Msun}{\mbox{M$_\odot$}}

\newcommand{\halos}{haloes}
\newcommand{\Halos}{Haloes}
\newcommand{\subhalos}{subhaloes}

\newlength{\figwidth}
\newlength{\figtable}
\newlength{\figtripple}
\newlength{\resplot}
\title[Influence of Halo Finders in Merger Trees]
{Sussing Merger Trees: the influence of the halo finder}
\author[Avila et al.]
{Santiago Avila$^{1,2}$\thanks{e-mail: santiago.avila@uam.es}, Alexander~Knebe$^1$, Frazer~R.~Pearce$^3$, Aurel~Schneider$^4$,
\newauthor
Chaichalit~Srisawat$^4$, Peter~A.~Thomas$^4$, Peter~Behroozi$^{5}$, Pascal~J.~Elahi$^6$, 
\newauthor
 Jiaxin~Han$^{7,8}$, Yao-Yuan~Mao$^{5}$, Julian~Onions$^3$, Vicente Rodriguez-Gomez$^{9}$, 
\newauthor
Dylan~Tweed$^{10,11}$
\\
  $^{1}$Departamento de F\'isica Te\'orica, M\'odulo C-15, Facultad de Ciencias, Universidad Aut\'onoma de Madrid, 28049 Cantoblanco, Madrid, Spain\\
  $^{2}$Instituto de F\'isica Te\'orica, UAM-CSIC, Universidad Autonoma de Madrid, 28049 Cantoblanco, Madrid, Spain\\
  $^{3}$School of Physics \& Astronomy, University of Nottingham, Nottingham, NG7 2RD, UK\\
  $^{4}$Department of Physics \& Astronomy, University of Sussex, Brighton, BN1 9QH, UK\\
  $^{5}$Space Telescope Science Institute, Baltimore, MD 21218, USA \\
  $^{6}$Sydney Institute for Astronomy, University of Sydney, Sydney NSW 2016, Australia\\
  $^{7}$Key Laboratory for Research in Galaxies and Cosmology, Shanghai Astronomical Observatory, 80 Nandan Road, Shanghai 200030, China \\
  $^{8}$Institute for Computational Cosmology, Department of Physics, Durham University, South Road, Durham DH1 3LE, UK\\
  $^{9}$Harvard-Smithsonian Center for Astrophysics, 60 Garden Street, Cambridge MA, 02138, USA\\
  $^{10}$Racah Institute of Physics, The Hebrew University, Jerusalem 91904, Israel \\ 
  $^{11}$Center for Astronomy and Astrophysics, Department of Physics and Astronomy,
Shanghai Jiao Tong University, 955 Jianchuan road, \\
Shanghai 200240, China\\
  }
\begin{document}
\date{\today}

\pagerange{\pageref{firstpage}--\pageref{lastpage}} \pubyear{2013}\volume{0000}

\maketitle

\label{firstpage}

\begin{abstract}
Merger tree codes are routinely used to follow the growth and merger
of dark matter \halos\ in simulations of cosmic structure formation.
Whereas in Srisawat et. al. we compared the trees built using a wide
variety of such codes here we study the influence of the underlying
halo catalogue upon the resulting trees.  We observe that the
specifics of halo finding itself greatly influences the constructed
merger trees.  We find that the choices made to define the halo mass
are of prime importance. For instance, amongst many potential options
different finders select self-bound objects or spherical regions of
defined overdensity, decide whether or not to include substructures
within the mass returned and vary in their initial particle selection.
The impact of these decisions is seen in tree length (the period of
time a particularly halo can be traced back through the simulation),
branching ratio (essentially the merger rate of \subhalos) and mass
evolution.  We therefore conclude that the choice of the underlying
halo finder is more relevant to the process of building merger trees
than the tree builder itself.  We also report on some built-in
features of specific merger tree codes that (sometimes) help to
improve the quality of the merger trees produced.
\end{abstract}


\section{Introduction} \label{sec:introduction}

The backbone of any semi-analytical model of galaxy formation is a merger tree of dark matter haloes. 
Some modern semi-analytical codes \citep{Croton06, Somerville08, Monaco07,
Henriques09, Benson12} rely on purely analytical forms
such as \citet{Press74} or Extended Press-Schechter \citep{Bond91}
--see \citet{Jiang13} for a recent comparison of such
methods--, while other codes take as input halo merger trees derived
from large numerical simulations (see \citet{RQP97,LC93} for the historical origin
of both approaches). 
Therefore, stable semi-analytic models require well-constructed and
physically realistic merger trees: \halos\ should not dramatically
change in mass or size, or jump in physical location from one step to
the next. There are two main steps required for the production of
merger trees from a $N$-Body simulation; firstly each output timeslice from a simulation needs to
be analysed to produce a halo catalogue, a step
performed by a halo finding algorithm. Secondly these halo catalogues
need to be linked together across snapshots by a tree building algorithm to construct a
merger tree. It is this final merger tree that is taken as input by a
semi-analytical model.

The properties of the merger trees built using a variety of different
methods has been addressed in the first ever comparison of tree
builders by \cite{Srisawat13}, a paper that emerged from our
\textsc{Sussing Merger Trees}
workshop.\footnote{http://popia.ft.uam.es/SussingMergerTrees} While we
observed that different tree building algorithms produce distinct
results, the influence of the underlying halo catalogue (the first
stage of the the two step process mentioned above) still remained
unanswered. This is nevertheless an important question as different
groups rely on their individual pipelines, which often includes their
own simulation software, halo finding method and tree construction
algorithm before the trees are fed to a semi-analytical model to
obtain galaxy catalogues.

In a series of comparisons of (sub-)halo finders
\citep[e.g.][]{Knebe11,Knebe_galaxies_2013,onions_2012,Onions13spin,streams},
which are all summarised in \citet{Knebe13}, we have seen that there
can be substantial variations in the halo properties depending on the
applied finder. This will certainly leave an imprint when using the
catalogue to construct merger trees. As a fixed input halo catalogue
was used for our first tree builder comparison the question remains;
to what extent are merger trees sensitive to the supplied halo
catalogue?

In this work we include both steps of the tree building process,
i.e. we will apply a set of different tree builders to a range of halo
catalogues constructed using a variety of object finders. Please note
that the underlying cosmological simulation remains identical in all
instances studied here. We are investigating how much of the scatter
in the resulting merger trees that form the input to semi-analytical
models stems from the tree building code and how much stems from the
halo finder. Or put differently, is a merger tree more affected by the
choice of the code used to generate the tree or the code used to
identify the dark matter \halos\ in the simulation?

In what follows, the input halo catalogues and the respective finders
they originate from will be presented in \Sec{sec:halos}. In
\Sec{sec:codes} we will then give a brief description of the merger
tree building codes. Our results will be reported in \Sec{sec:geom} and
\Sec{sec:mass}. We close with discussion and our conclusions in
\Sec{sec:disc}.

\section{Input halo catalogues} \label{sec:halos}

The halo catalogues used for this paper are extracted from 62
snapshots of a cosmological dark-matter-only simulation undertaken
using the \textsc{Gadget-3} $N$-body code
\citep{springel_cosmological_2005} with initial conditions drawn from
the WMAP-7 cosmology \citep{komatsu_etal_2011}.  We use $270^3$
particles in a box of comoving width $62.5$ $h^{-1}$ Mpc/h, with a
dark-matter particle mass of $m_p=9.31\times10^8h^{-1}$\Msun. 
We use 62 snapshots (000,\ldots,061) evenly spaced in $log\, a$
from reshift 50 to redshift 0.

While in previous comparison projects \citep[e.g.][]{Knebe11,Knebe13,onions_2012}
we forced the same mass definition (or even used a common
post-processing pipeline to assure this), we did not request any such
thing this time, i.e. every halo finder was allowed to use its own
mass definition.

On the one hand, \ahf\ and \rockstar\ define a \textit{spherically
truncated} mass through
\begin{equation}
\label{eq:virialradius}  M_{\rm ref}(<R_{\rm ref}) = \Delta_{\rm ref} \times \rho_{\rm ref} \times \frac{4\pi}{3} R_{\rm ref}^3 \ ,
\end{equation}
\noindent
adopting the values $\Delta_{\rm ref}=200$ and $\rho_{\rm
ref}=\rho_{\rm crit}$ (we will call this mass $M_{200c}$) and iteratively removing particles not bound to the structure.
On the other hand, \hbthalo\ and \subfind\ return \textit{arbitrarily
shaped} self-bound objects based upon initial Friends-of-Friends (FoF)
groups, assigning them the mass of \textit{all} (i.e. no spherical
truncation) particles gravitationally bound to the halo.

Furthermore, some halo finders include the mass of any bound
substructures in the main halo mass whereas others do not include the
mass of any bound substructures. Technically, finders for which
particles can only belong to one halo are termed \textit{exclusive}
while finders for which particles can belong to more than one halo are
termed \textit{inclusive}. As substructures can typically account for
10\% of the halo mass this choice alone can make a substantial
difference to the halo mass function.

Given these definitions we can now describe the general properties of
the halo finders applied to the data:

\begin{itemize}
        \item \ahf\ \citep{gill_evolution_2004,knollmann_ahf:_2009} is
a configuration-space Spherical Overdensity (SO) adaptive mesh
finder. It returns inclusive gravitationally bound \halos\ and
\subhalos\ spherically truncated at $R_{200c}$ (thus, the mass returned
is $M_{200c}$).
        \item \hbthalo\ \citep{han_etal_2012} is a tracking algorithm
working in the time domain that follows structures from one timestep
to the next. It returns exclusive arbitrarily shaped gravitationally
bound objects. It uses FoF groups for the initial
particle collection.
        \item \rockstar\ \citep{Behroozi_2013} is a phase-space halo
finder. A peculiarity of this code is that --unlike \ahf, \hbthalo\ and
\subfind-- the mass returned for a halo does not correspond to the sum
of the mass of the particles listed as belonging to it.  While it uses
the same mass definition as \ahf\ (inclusive bound $M_{200c}$ mass),
the particle membership list of the halo is exclusive and is made up
of particles close in phase-space to the halo centre.
        \item \subfind\ \citep{subfind_2001} is a configuration-space
finder using FoF groups as a starting point which are subsequently
searched for \subhalos. It returns arbitrarily shaped exclusive
self-bound main \halos, and arbitrarily shaped self-bound \subhalos\
that are truncated at the isodensity contour that is defined by the
density saddle point between the subhalo and the main halo.
\end{itemize}

\begin{figure}
 \includegraphics[height=1\linewidth,angle=270]{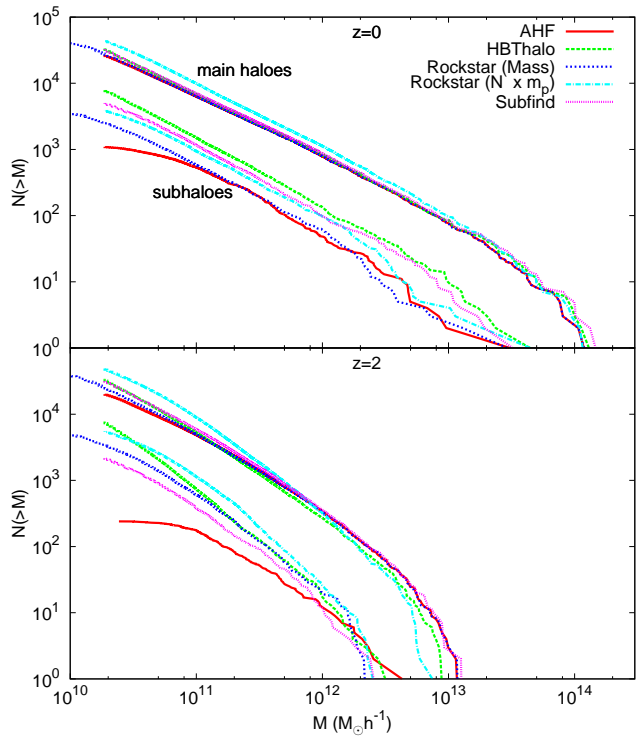}
\caption{Cumulative mass functions at redshift $z=0$ (upper panel) and
$z=2$ (lower panel) for the four halo finders. There are two lines for
\rockstar\ corresponding to the two mass definitions discussed in the
text: one corresponding to $M_{200c}$ (\textit{Mass}) and one based upon the particle list 
($N \times m_p$, being $N$ the number of particles and $m_p$ the particle mass).
The upper set of curves in each panel is based upon main \halos\
whereas the lower set of curves in each panel refers only to \subhalos.}
\label{fig:massfunc}
\end{figure}

To give an impression of the differences in the halo catalogues, we
present in \Fig{fig:massfunc} the cumulative mass function for the
four halo finders at redshift $z=0$ (upper panel) and $z=2$ (lower
panel); we further separate \subhalos\ from main \halos\ and present
their cumulative mass spectrum in the upper and lower set of curves of each panel,
respectively. We have set a threshold of 20 particles (equivalent to
$M = 20 m_p = 1.86\times 10^{10} h^{-1}M_{\sun}$) for \halos\ to be
considered.  In order to highlight the peculiarity of \rockstar\ (for
which the returned mass does not correspond to the sum of the mass of
the particle membership) we have plotted two lines for \rockstar: one
based upon summing individual particle masses (cyan dash-dotted) and one with the
mass $M_{200c}$ as returned by \rockstar\ (blue dotted, extending to masses below the 20
particle threshold).  Given that some tree builders only use particle
membership information for a halo whereas others combine this with a
table of global properties (including halo mass), this choice of mass
definition will also contribute to the differences in the final trees.

We find that other than for the largest 100 main \halos\ the different
mass definitions make little difference unless the mass taken from the
returned \rockstar\ particle membership is used. This mass is
systematically higher that the other estimates (and \rockstar's own
returned mass). The differences in mass for main \halos\ are slightly
more pronounced at $z=2$.

For \subhalos\ there are noticeably different mass functions: \ahf\ is
incomplete at the low-mass end, with a trend that appears
to worsen as the redshift increases\footnote{We confirm (though not explicitly 
shown here) that a more restrictive parameter set for \ahf\ leads to the recovery of the missing 
low-mass subhaloes at high redshift. As already shown by \citet{knollmann_ahf:_2009} 
(Fig.5 in there) there is a direct dependence of the applied refinement threshold 
used by \ahf\ to construct its mesh hierarchy (upon which haloes are based) 
to the number of low-mass objects found.}.  However, despite generally
finding more \subhalos\ the other finders do not appear to have
converged to a common set.  Part of this relates to the rather
ambiguous definition of subhalo mass: whereas for main \halos\ it
simply appears to be a matter of choice for $\Delta_{\rm ref}$ and
$\rho_{\rm ref}$ (or some other well-defined criterion for
virialisation/boundness/linkage), \subhalos\ -- due to the embedding
within the inhomogeneous background of the host -- cannot easily
follow any such rule.  Again, each finder has been allowed to pick its
favourite definition for subhalo mass. But please note that the
variations seen here are not the prime focus of this study; they
should nevertheless be taken into account when interpreting the
results presented and discussed below. Further, the scatter in subhalo
mass functions seen in previous comparisons was much reduced due to
the use of a common post-processing pipeline that ensured a unique
subhalo mass definition \citep[]{onions_2012,Onions13spin,Knebe13}.

All these differences should and will certainly leave an imprint and
be reflected in the outcome when building merger trees.

\section{Merger Tree Building Codes} \label{sec:codes}
The participating merger tree building codes have been extensively
described and classified in the original comparison paper
\citep{Srisawat13}.  But as not all merger tree builders from the original
comparison engaged in this particular study and for completeness we briefly
describe the participating tree building codes here.

\begin{figure*}
  \centering
  \vspace*{0cm}\hspace*{-4cm}
  \includegraphics[width=0.6\linewidth]{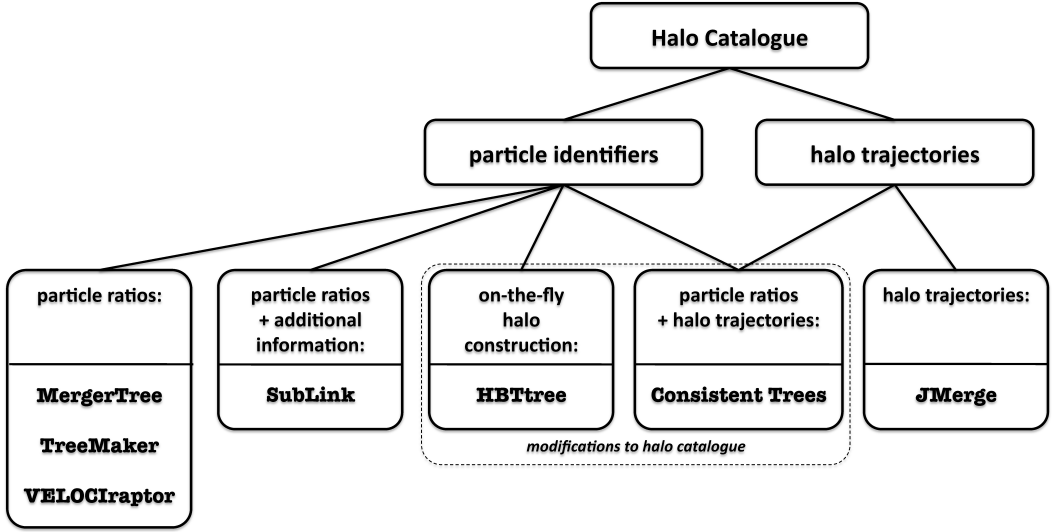}
  \caption{A summary of the main features and requirements of the
    different merger tree algorithms.
For details see the individual descriptions in the text.}
\label{fig:overview}
\end{figure*}

As a lot of the underlying methodology is similar across the various
codes used here we have tried to capture the main features and
requirements in \Fig{fig:overview}. We first categorise tree builders
into either using halo trajectories (\jmerger, and \consistenttree) or
individual particle identifiers (together with possibly some
additional information; all remaining tree builders). \consistenttree\
is the only method that utilises both types of approach. \hbt\
constructs halo catalogues and merger trees at the same time as it is
a tracking finder that follows structures in time.  A cautionary note
regarding \hbt: it can be applied both as a halo finder or a tree
builder and includes elements of both so we will always specify
whether we refer to one or the other by appending `halo' or `tree', as
necessary.

The codes themselves are best portrayed as follows:

\begin{itemize}
        \item \consistenttree\ forms part of the \rockstar\
        package. It gravitationally evolves positions and velocities
        of \halos\ between timesteps, making use of information from
        surrounding snapshots to correct missing or extraneous \halos\
        in individual snapshots \citep{Behroozi_etal_2013}.
        \item \hbttree\ is built into the halo finder \hbt. It
        identifies and tracks objects at the same time using particle
        membership information to follow objects between output times.
        \item \jmerger\ only uses halo positions and velocities to
        construct connections between snapshots, i.e. \halos\ are moved
        backwards/forward in time to identify matches that comply with
        a pre-selected thresholds for mass and position changes.
        \item \mergertree\ forms part of the \ahf\ package and
        cross-correlates particle IDs between snapshots.
        \item \sublink\ tracks particle IDs in a weighted fashion,
        giving priority to the innermost parts of \subhalos\ and
        allowing branches to skip one snapshot if an object disappears.
        \item \treemaker\ consists of cross-comparing (sub)\halos\ from
        two consecutive output times by tracing their exclusive sets
        of particles.
        \item \velociraptor\ is part of the \velociraptor/\stf\ package
        and cross-correlates particle IDs from two or more structure
        catalogues.
\end{itemize}

Two codes were allowed to modify the original catalogue:
\consistenttree\ and \hbttree. \consistenttree\ adds \halos\ when it
considers they are missing: i.e., the halo was found both at an earlier
and at a later snapshot. \consistenttree\ also removes \halos\ when it
considers them to be numerical fluctuations: i.e., the halo does not
have a descendant and both merger and tidal annihilation are unlikely
due to the distance to other \halos.  \hbttree\ for \textit{external}
halo finders (i.e. halo catalogues not generated by its own inbuilt
routine) takes the main halo catalogue and reconstructs the
substructure. This produces an exclusive halo catalogue in which the
properties of the main \halos\ may also have changed.

The participants were asked to build merger trees starting from our input
halo catalogues described in Section~\ref{sec:halos}, in the same
way as was done for the original comparison presented in
\citet{Srisawat13}.

\section{Geometry of trees}\label{sec:geom}

In this section we study the geometry and structure of merger trees
and the resulting evolution of dark matter \halos.  This includes the
length of the tree (\Sec{sec:length}) and the tree branching ratio
(\Sec{sec:branches}). We further show graphically how halo finders and
tree builders work differently, to illustrate the features found in
the comparison.

\begin{figure*}
  \centering
  \includegraphics[height=0.5\linewidth,angle=270]{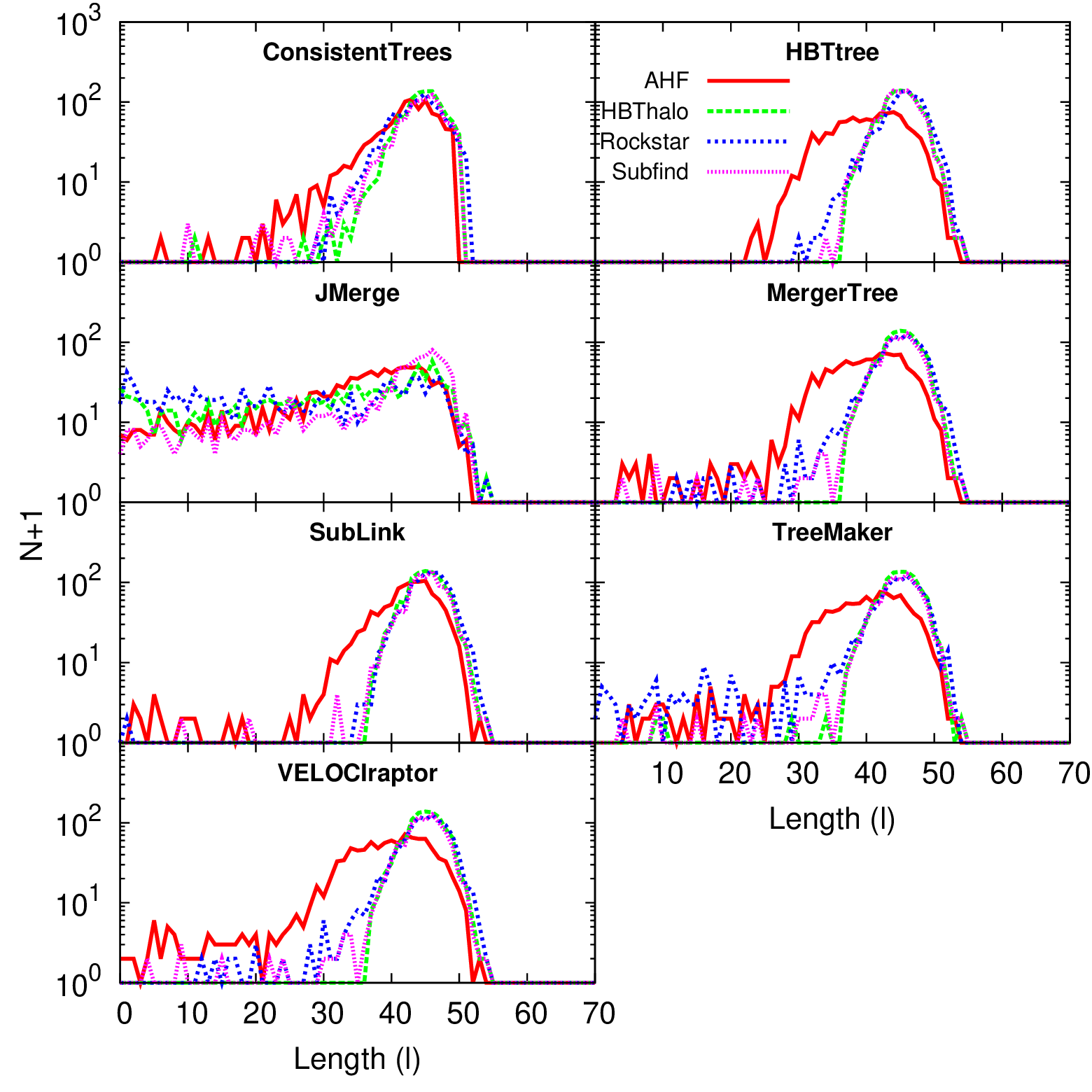}\includegraphics[height=0.5\linewidth,angle=270]{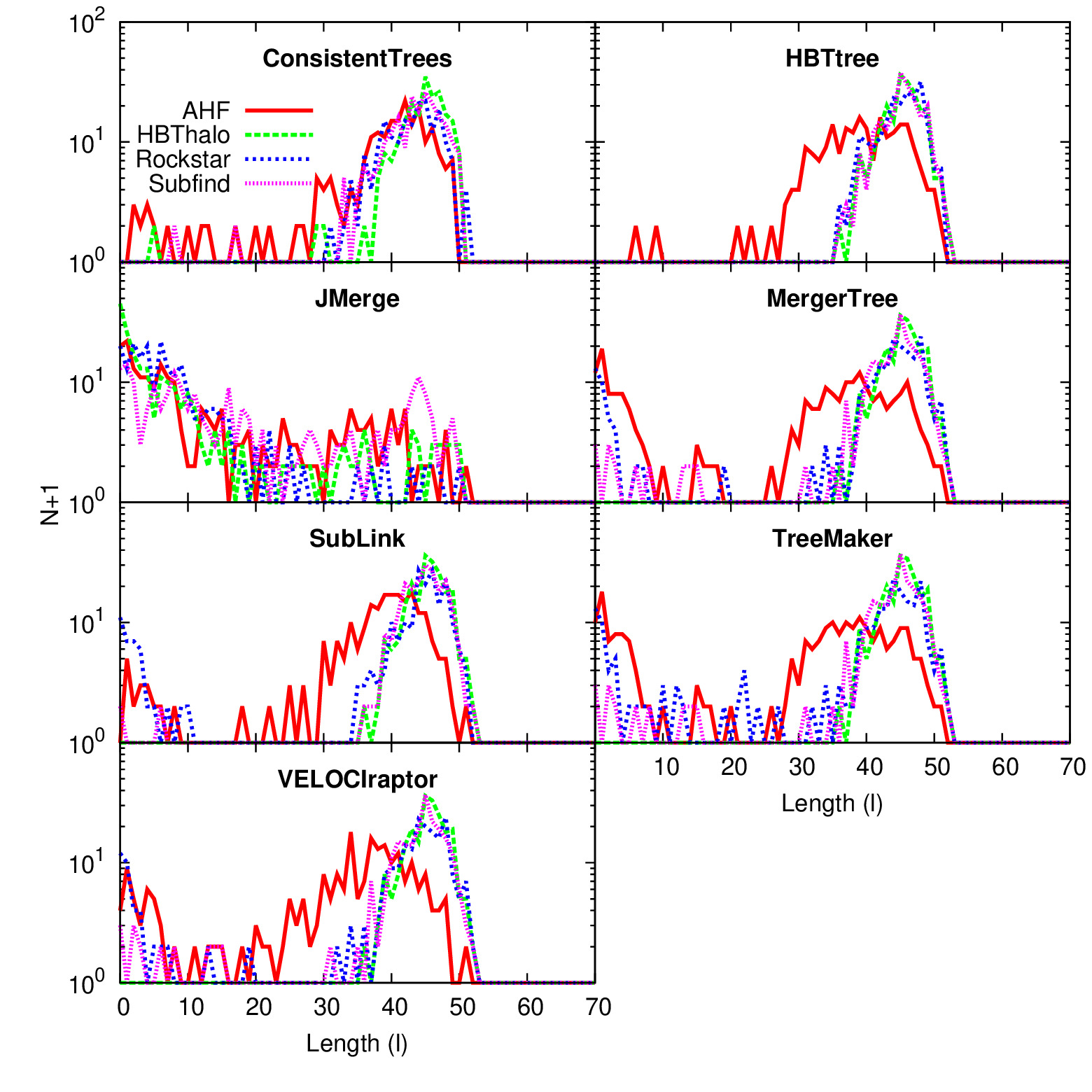}
  \caption{Histogram of the length of the main branch.
    The length $l$ is defined as the number of snapshots a halo can be
    traced back through from $z=0$.  The left group of panels show the
    $1000$ most massive main \halos.  The right group of panels show
    the $200$ most massive \subhalos.  These number selections are
    equivalent to the mass cuts shown in \Tab{tab:masscuts}.
    Different panels contain results from different tree building
    methods (as indicated), while within each panel there is one line
    for each halo finder (as marked in the legend).  }
\label{length}
\end{figure*}

\subsection{Length of main branches} \label{sec:length}

One of the conceptually simplest properties of a tree is the length of
the main branch.\footnote{We use the terminology introduced in Sec.~2
  of \citet{Srisawat13} throughout this paper.}  It measures how far
back a halo can be traced in time -- starting in our case at $z=0$.
This property not only relies on the performance of the halo finder
and its ability to identify \halos\ throughout cosmic history, but
also on the tree builder correctly matching the same halo between
snapshots. \citet{Srisawat13} found that the different tree building
methods produced a variety of main branch lengths, ascribing
some of the features to halo finder flaws. We shall verify this now.

\Fig{length} shows a histogram of the main branch length $l$, defined
as the number of snapshots a halo main branch extends backwards in
time from snapshot $61$ ($z=0$) to snapshot $61-l$.  This is roughly
equivalent to an age, given that the last 50 snapshots are separated
uniformly in expansion factor, $a = 1/1+z$. On the left, we selected
the $1000$ most massive main \halos, whereas on the right we show the
results for the $200$ most massive \subhalos.  The main halo
population coincides from one halo catalogue to another in at least
$85\%$ of the objects. The subhalo population is more complicated and,
in some cases, they only agree in $15\%$ of the objects from one
finder to another.  However, if we focus on comparing \ahf\ with
\rockstar\ or \hbthalo\ with \subfind, we find a better agreement
between catalogues, rising to $\sim 95\%$ for main \halos\ and
$\sim70\%$ for \subhalos.  Due to these differences, the applied
number threshold translates to mass thresholds $M_{th}$ that are
different from finder to finder (see also \Fig{fig:massfunc}); we
therefore list the corresponding values in
\Tab{tab:masscuts}. Furthermore, when using \hbttree, the individual
masses of the \halos\ can change and so does the mass threshold.  In
what follows we will consistently use these mass thresholds, even at
higher redshift.

  \begin{table}
      \begin{tabular}{lrrrr@{}}
       & \ahf\ & \hbthalo\ & \rockstar\ & \subfind\ \\
      \hline
	$M_{th}^{\rm main}$ & $7.93$& $8.25$ & $7.90$ & $9.61$ \\
	$M_{th,\rm HBT}^{\rm main}$ & $7.52$ & $8.25$ & $10.64$ & $8.30$ \\
	$M_{th}^{\rm sub}$ & $3.09$ & $6.91$ & $3.00$ & $5.30$ \\
	$M_{th,\rm HBT}^{\rm sub}$ & $2.75$ & $6.91$ & $2.68$ & $5.90$ \\
      \hline
      \end{tabular}

      \caption{Mass threshold in units of $10^{11} h^{-1}M_{\sun}$
        needed to select at $z=0$ the $1000$ most massive main \halos\
        (rows 1 and 2) and the $200$ most massive \subhalos\ (rows 3
        and 4) for different halo finders (columns). Odd rows show the
        threshold for a general tree builder, whereas even rows show
        the threshold for \hbttree.}
  \label{tab:masscuts}
\end{table}

As expected by the hierarchical structure formation scenario induced
by cold dark matter, most large mass objects can be traced back to
high redshift.  This is not surprising and has already been reported
in our previous comparison, but here we can appreciate that this
result depends on the choice of the halo finder and we will elaborate
on this below.

As a general observation, for both main \halos\ and \subhalos, it is
apparent that \hbthalo\ leads to the best results: nearly all massive
\halos\ are found and followed from an early origin.  We attribute
this to the fact that by its very nature as a tracking finder \hbthalo\ is
designed with the intention of building a merger tree in mind.
\subfind\ tends to give similar results but with occasional early
truncation. These truncations become more pronounced for \ahf\ and
\rockstar.  Further, \ahf\ tends to terminate each tree slightly
earlier, even if it was well followed back in time, because of the
incompleteness at low mass end (\Fig{fig:massfunc}). For \ahf\ missing
low-mass objects at high redshift cannot be the small progenitors of
the high-mass low-redshift objects followed in \Fig{length}.

Differences between \subhalos\ and main \halos\ are also
apparent. First, the subhalo curves in general appear more noisy, in
part due to having fewer objects, but also because they are always
placed in a more complicated environment which enhances the
stochasticity. The difficulty in following \subhalos\ then causes more
cases with low $l$, especially for \ahf\ and \rockstar.  
One could naively think his excess of low l subhaloes for \ahf\ and \rockstar\ 
could be the result of a much smaller $M_{th}^{sub}$ threshold (see \Tab{tab:masscuts}). 
However, we verified that using the same threshold for all catalogues, 
only mitigates that difference without completely erasing it.


Subhalo finding becomes especially difficult as the subhalo approaches the
centre of the host halo, as has been shown in Fig.4 of
\citet{Sub_detect} and Fig.7 of \citet{onions_2012}.  In particular,
\subfind\ underestimates the mass of \subhalos\ close to the centre of
their host halo.  Given that the $200$ most massive \subhalos\ are not
the same for all finders, the \subhalos\ selected for \subfind\ tend
to be further from the host halo centres (not explicitly shown here),
and therefore they are easier to trace.  \ahf\ and especially
\rockstar\ find many (massive) \subhalos\ near the centre but, due to
the difficulties in that region, a fraction of them cannot be provided
with a credible progenitor in an earlier snapshot, resulting in
early tree termination.  Finally, the \hbthalo\ selection is composed
of \subhalos\ at short, medium and large distances from the host halo
centre but, by construction, they are always required to be trackable.

On the tree builder side, \jmerger\ allows \halos\ to only shrink their mass by a
factor of up to 0.7 and to grow by a factor of up to 4 in one
snapshot, and it estimates their trajectories from global quantities
(\Sec{sec:codes}).  This artificially truncates main branches too
early for massive objects when it loses track of \halos.  This effect
is enhanced for \subhalos, whose trajectories are difficult to
estimate due to the non-linear environment and the fact that their
mass is more likely to grow or shrink abruptly
(\Sec{sec:mass}). \consistenttree\ and \hbttree\ essentially eliminate
the low-$l$ cases for nearly all the \halos\ (and \subhalos). This is
due to their freedom to modify the catalogue in such a way as to avoid
exactly these occurrences.

\begin{figure*}
  \centering
  \includegraphics[height=0.5\linewidth,angle=270]{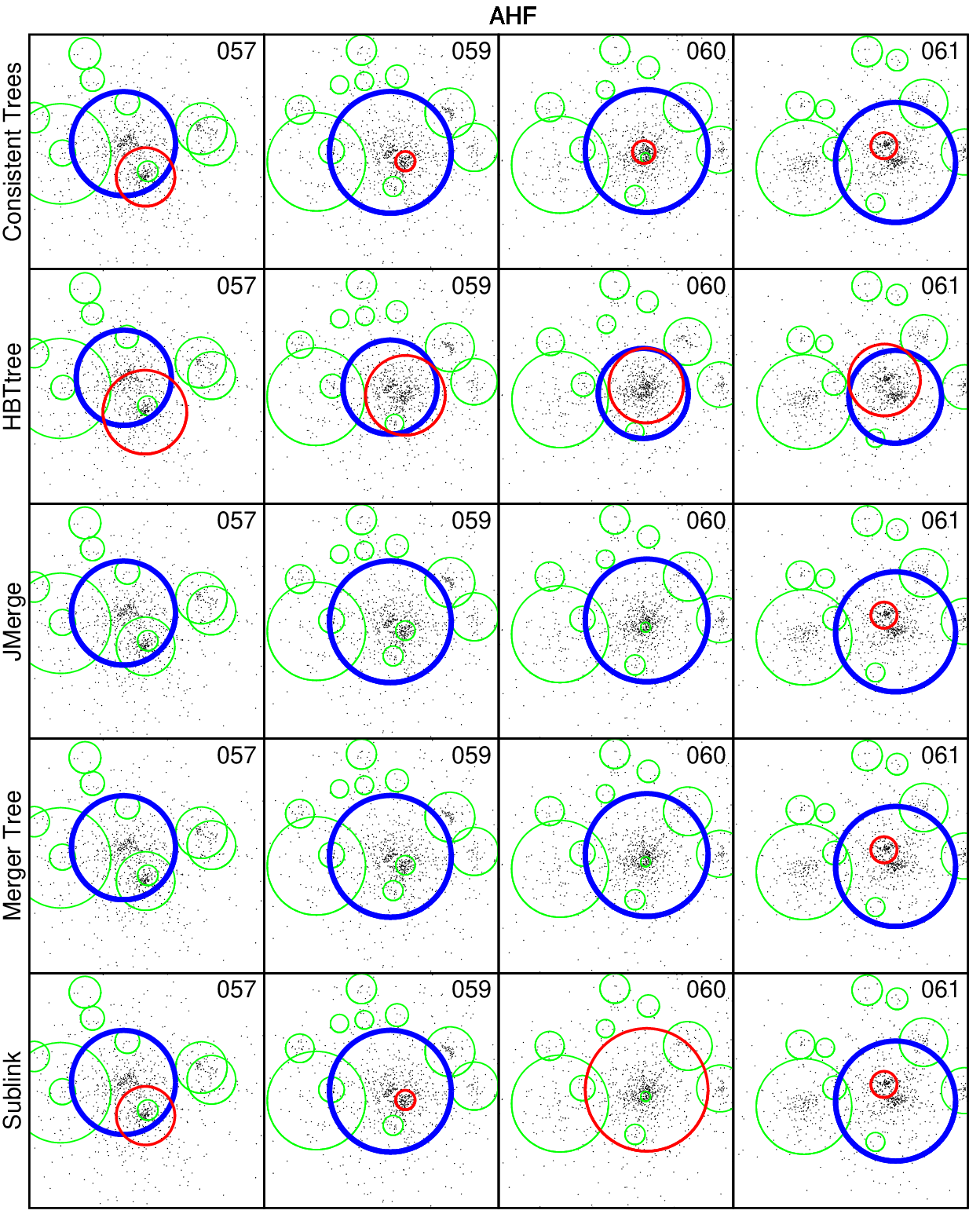}\includegraphics[height=0.5\linewidth,angle=270]{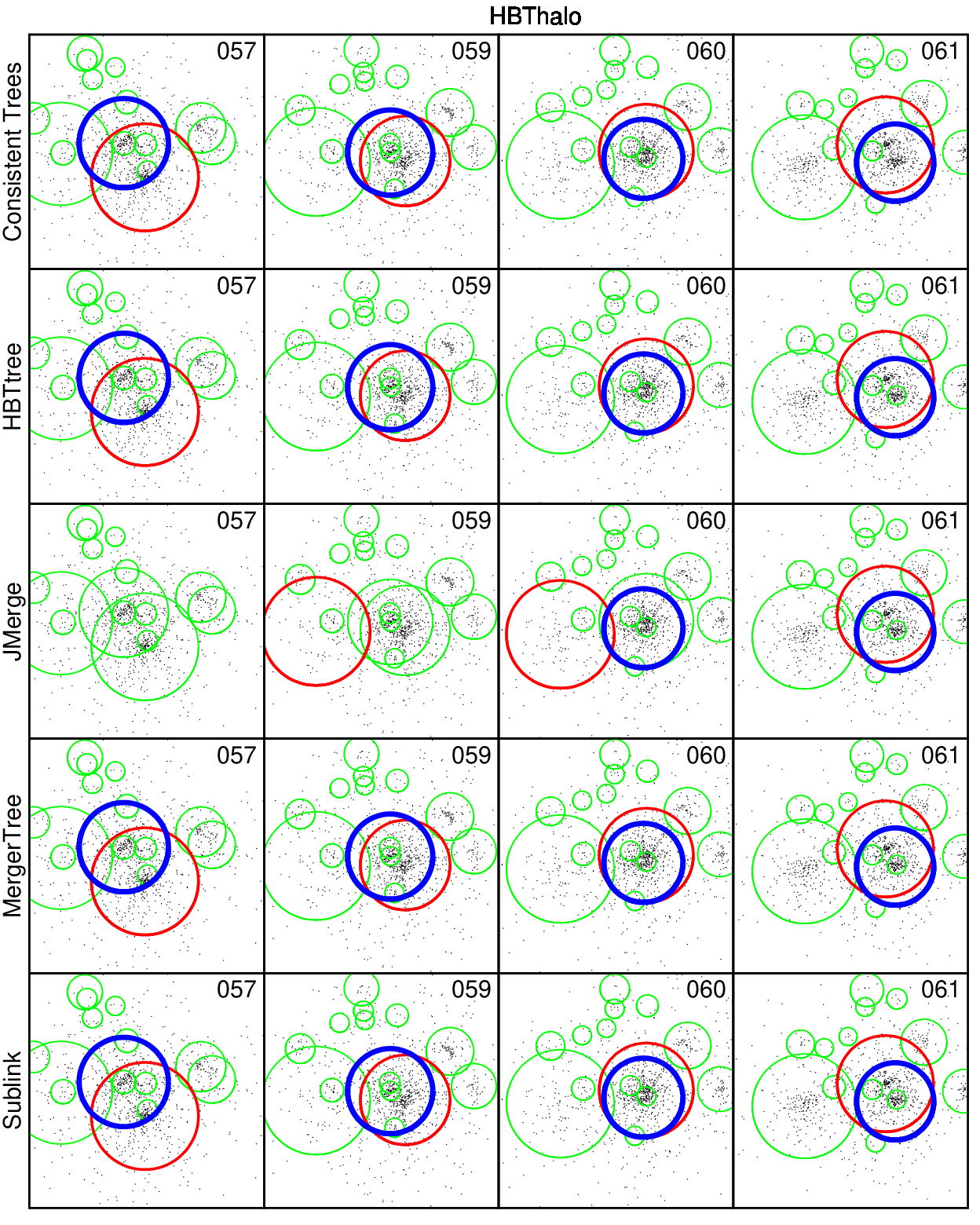}
  \includegraphics[height=0.5\linewidth,angle=270]{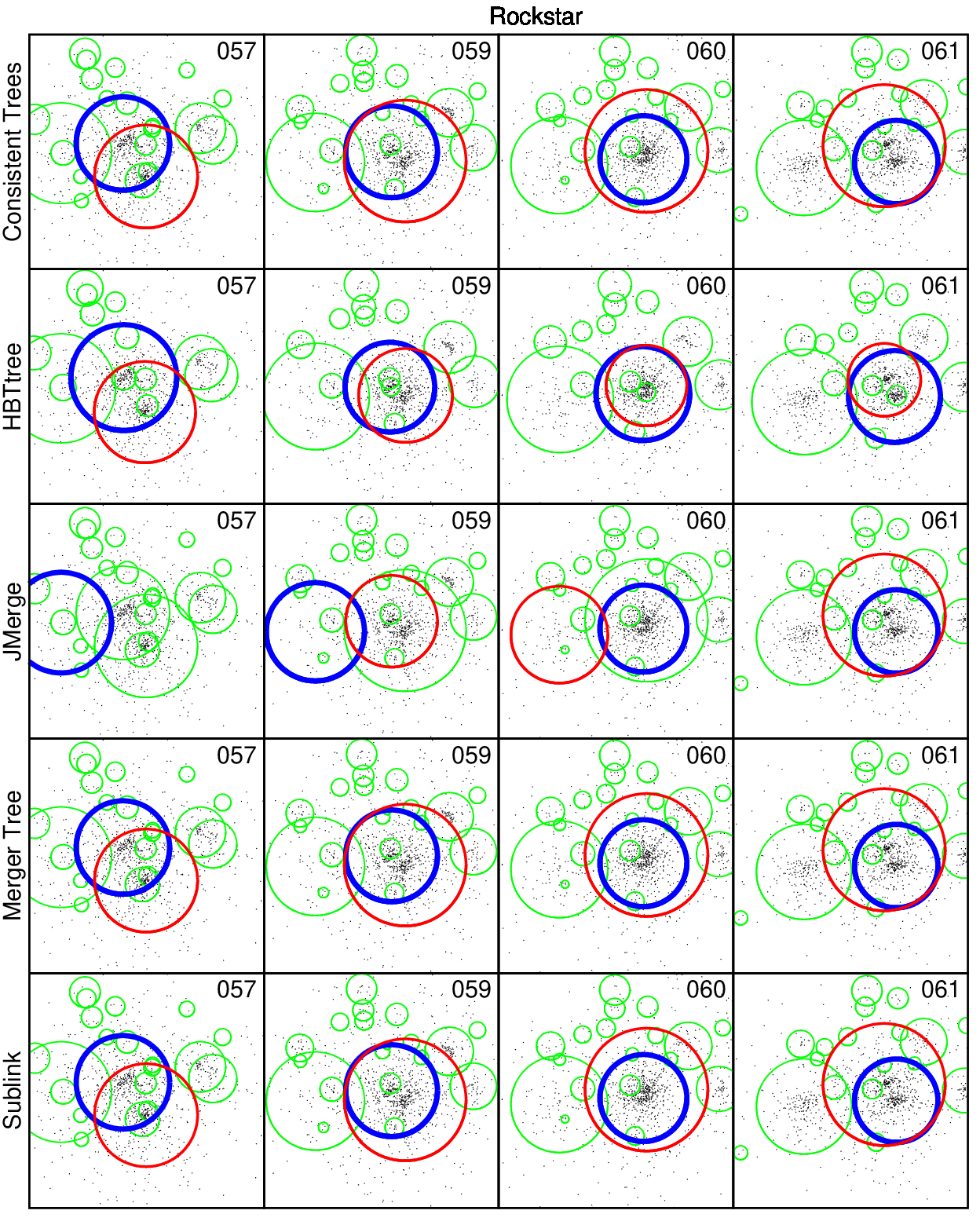}\includegraphics[height=0.5\linewidth,angle=270]{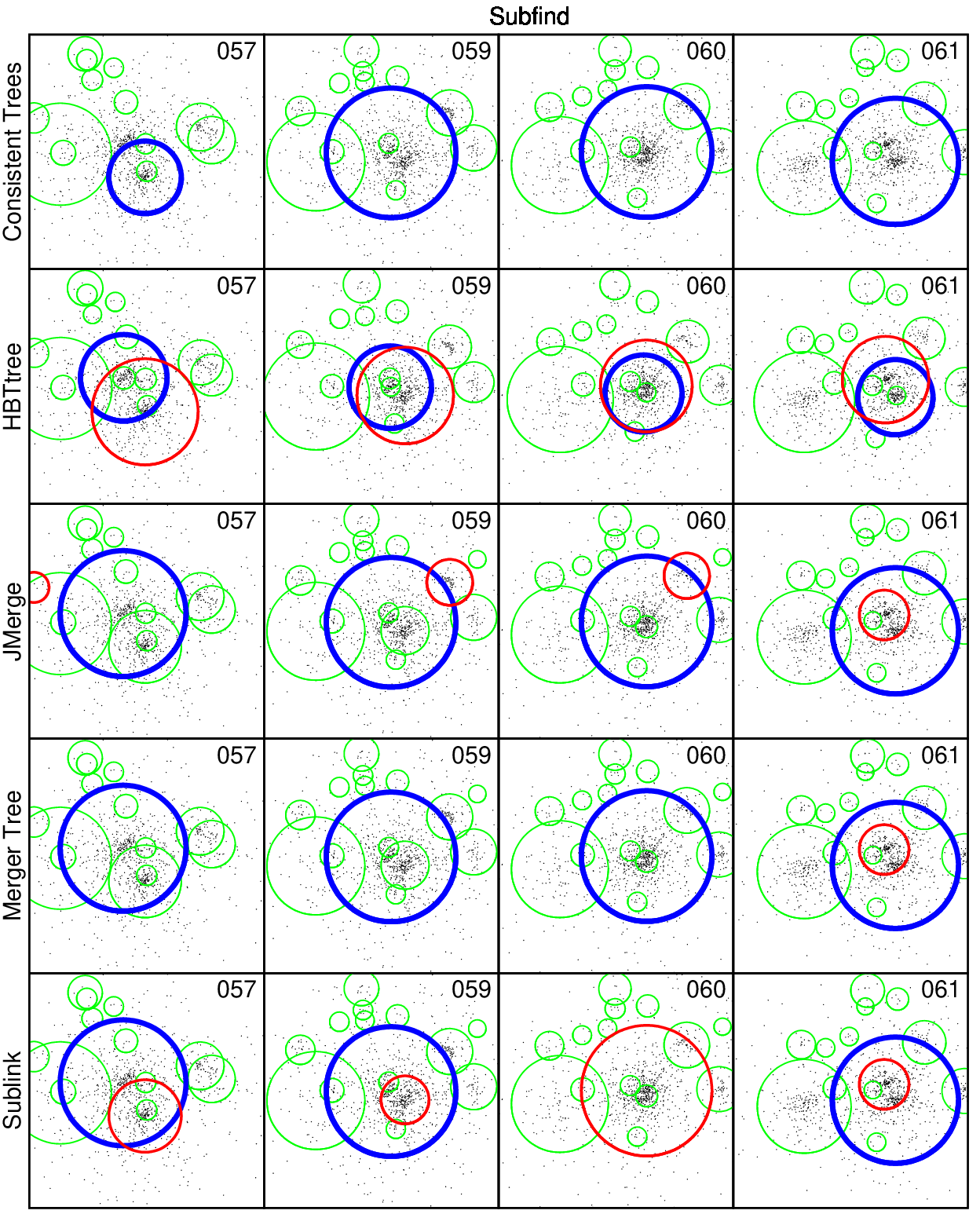}
  \caption{Projected image of a $1.2$ Mpc/h-side cube from the $N$-Body
    simulation. \Halos\ are represented by circles of radius
    corresponding to $R_{200c}$.  This is an example of a merger
    between two \halos\ that are found at $z=0$ (snapshot 061) and
    linked across snapshots by the tree builders: the blue (thickest line) and red (intermediate thick line)
    colours represent the two trees. Other haloes found are represented in green (thinest line).
    Each subfigure presents a single
    halo finder, with each row representing the indicated tree
    builder. In each row time evolves from left to right, with each
    cell a different snapshot (labeled at the top right corner of the
    cell).  Note that the missing tree builders all gave the same
    result as \mergertree .}
\label{case}
\end{figure*}

In order to better illustrate the factors that influence the main
branch length, $l$, we present in \Fig{case} a graphical
representation of the performance of the various halo finders and tree
builders. The figure shows a projected $1.2$ Mpc/h-side cube extracted
from the $N$-Body simulation with the particles (dots), the \halos\
found (circles), and the construction of two trees (specific thickness and colour).  
This slice shows two \halos\ of similar size passing through
each other in the process of a merger \citep[the same merger as shown
in Fig.4 of ][]{Srisawat13}. These two \halos\ are identified at $z=0$
by a thick blue line and an intermediate thickness red line, and then traced back 
by the merger tree.  For this example  \mergertree, \treemaker\ and
\velociraptor\ gave identical results and so we only show the
\mergertree\ result.

We find a wide variety of situations: in some cases every halo is
correctly traced (e.g. \consistenttree\ with \ahf) but in
others the tracing fails (e.g. \jmerger\ with \ahf).  In the success
or failure of the tracing the influence of both the halo finder and
the tree builder are important. The effect of the tree builder was
already reported in \cite{Srisawat13}, so here we focus on emphasising
the dependence on the halo finder:

\begin{itemize}
\item \ahf\ considers one of the merging haloes to be the main
  halo (blue) and the other to be a subhalo (red).  In snapshot 060
  the subhalo found is quite small, so that most of the tree building
  codes do not link it with the (much larger) halo in the next
  snapshot (061). In simple codes (\jmerger, \mergertree... ) this
  leads to an artificial truncation of the tree. \consistenttree\
  artificially adds one halo to snapshot 060 to replace the small
  subhalo whereas \sublink\ jumps snapshot 060 for this object. In
  this way both codes continue the tree. \hbttree\ recomputes the
  substructure, creating a more trackable subhalo.

\item \hbthalo\ is able to identify at snapshot 060 two big and well
  defined \halos\ of almost the same size (only possible for exclusive
  halo catalogues).  This is due to the tracking nature of the finder and
  ensures the correct follow-up by most tree builders.  Only
  \jmerger\ encounters problems due to the non-smooth trajectories of
  the \halos.

\item \rockstar\ uses phase-space information so that even when the
  \halos\ are overlapping (snapshot 060) it is able to distinguish
  them by their velocities.  This allows almost all tree codes
  (besides \jmerger) to follow the evolution of the \halos.

\item \subfind\ gives similar problems to \ahf: the subhalo at snapshot 
  060 is too small to be considered a credible progenitor. For this 
  catalogue, \consistenttree\ is not able to deal with it and completely
  removes the red tree. \hbttree\ patches over that problem the usual way
  while \jmerger\ 
  associates the halo to a progenitor incorrectly and \mergertree\ truncates the 
  tree. \sublink, by omitting snapshot 060, is able to follow the history correcty.
 
\end{itemize}

This example neatly illustrates the difficulties that arise when
dealing with \subhalos. However, the left panel of \Fig{length} tells
us that there are also situations in which the main halo branch is
truncated. We studied several of these cases and found two main types:
in the first type the main halo lies in the vicinity of a bigger halo,
and is likely to enter it and become a subhalo within the next few
snapshots. In this case the problems encountered are similar to those
illustrated in the subhalo example above, but here the infalling halo
is still classified as a main halo at $z=0$. The other type occurs
when at some point the halo was wrongly associated to some other
smaller halo as happened with the red halo in \Fig{length} for the
combination \jmerger -\hbthalo. In this case the incorrect halo
assignment never gets corrected and typically the much smaller halo
has a much shorter prior history.

Already at this stage of the analysis we can draw some conclusions
from this subsection:

\begin{itemize}
\item In general, the influence of the halo finder is at least as (if
  not more) important than the tree building algorithm.
\item Main \halos\ are easier to trace.
\item The way the halo finder deals with substructure is crucial
  for merger trees.
\item Tree building \textit{tricks} such as the creation of artificial
  \halos\ or omitting snapshots help in some cases, but are not
  infallible.
\item \ahf\ and \rockstar\ catalogues lead to earlier tree truncation
  for most tree builders. This is especially true for \subhalos, because they try
  to find \subhalos\ close to the host halo centre and are not
  able to provide them with credible progenitors.
\item \subfind\ tends to find more subhaloes in the outer regions of the host, which are easier to track.
\item \hbt\ appears to be very well designed to not truncate a tree
  too early, both as a halo finder and as a tree builder (as seen in
  \Fig{length} and \ref{case}).
\item \consistenttree\ also stands out in avoiding low-$l$ cases
  (\Fig{length}).
\item \jmerger\ faces problems in complex environments.
\end{itemize}

\subsection{Branching ratio} \label{sec:branches}

\begin{figure}
 \includegraphics[height=\figwidth,angle=270]{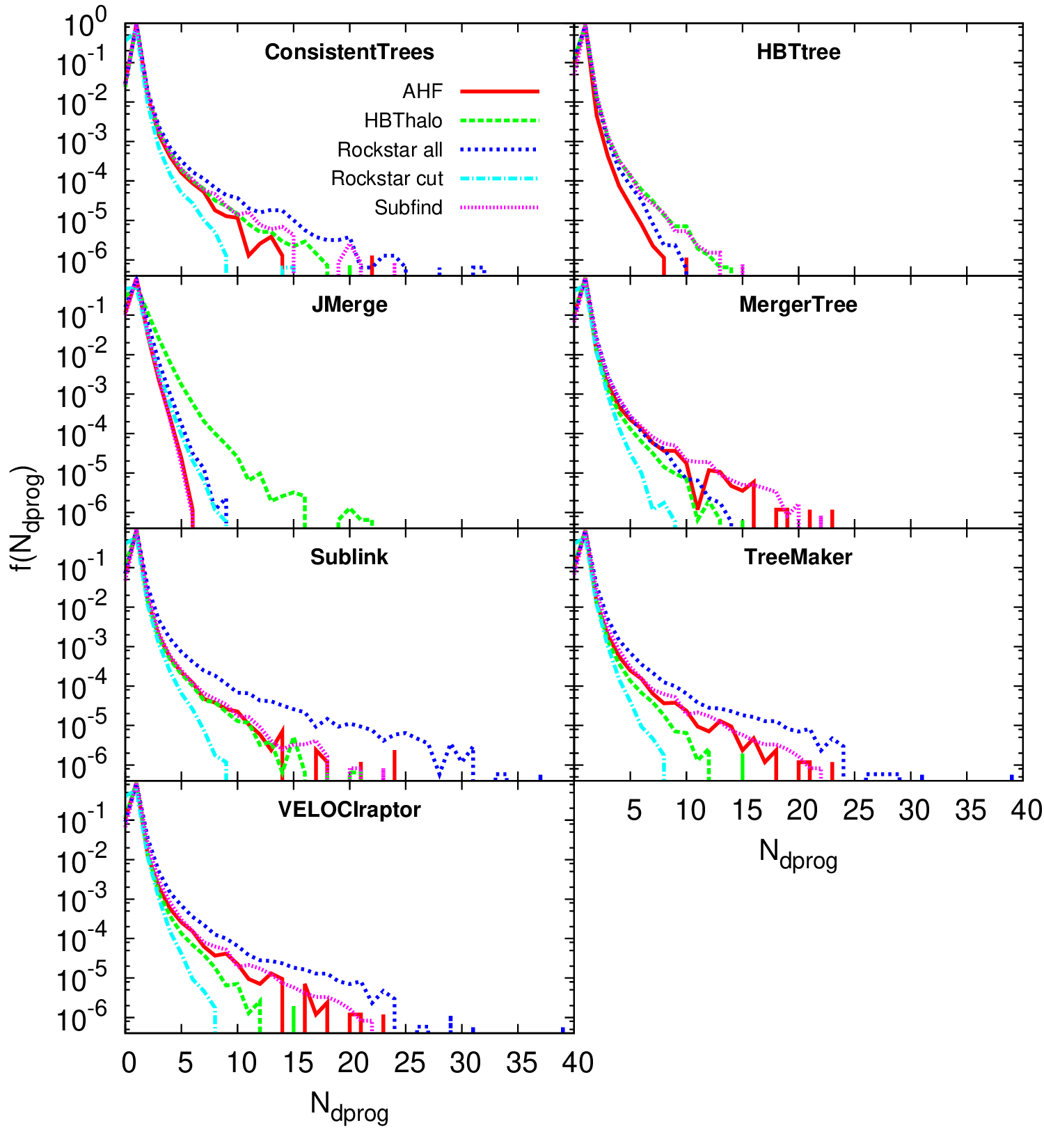}
 \caption{Normalised histograms of the number of direct progenitors
   $N_{\rm dprog}$ for all \halos\ from $z=0$ to $z=2$ (snapshots from
   $061$ to $031$).  Each panel corresponds to a single tree building
   method, within each panel each line represents a halo catalogue as
   indicated. For \rockstar\ we show two lines, one with all the
   \halos\ ('\rockstar\ all') and one where \halos\ with mass lower
   than $20\ m_p$ were removed ('\rockstar\ cut').}
\label{fig:Nprog}
\end{figure}

\begin{figure*}
  \centering
  \includegraphics[height=0.85\linewidth,angle=270]{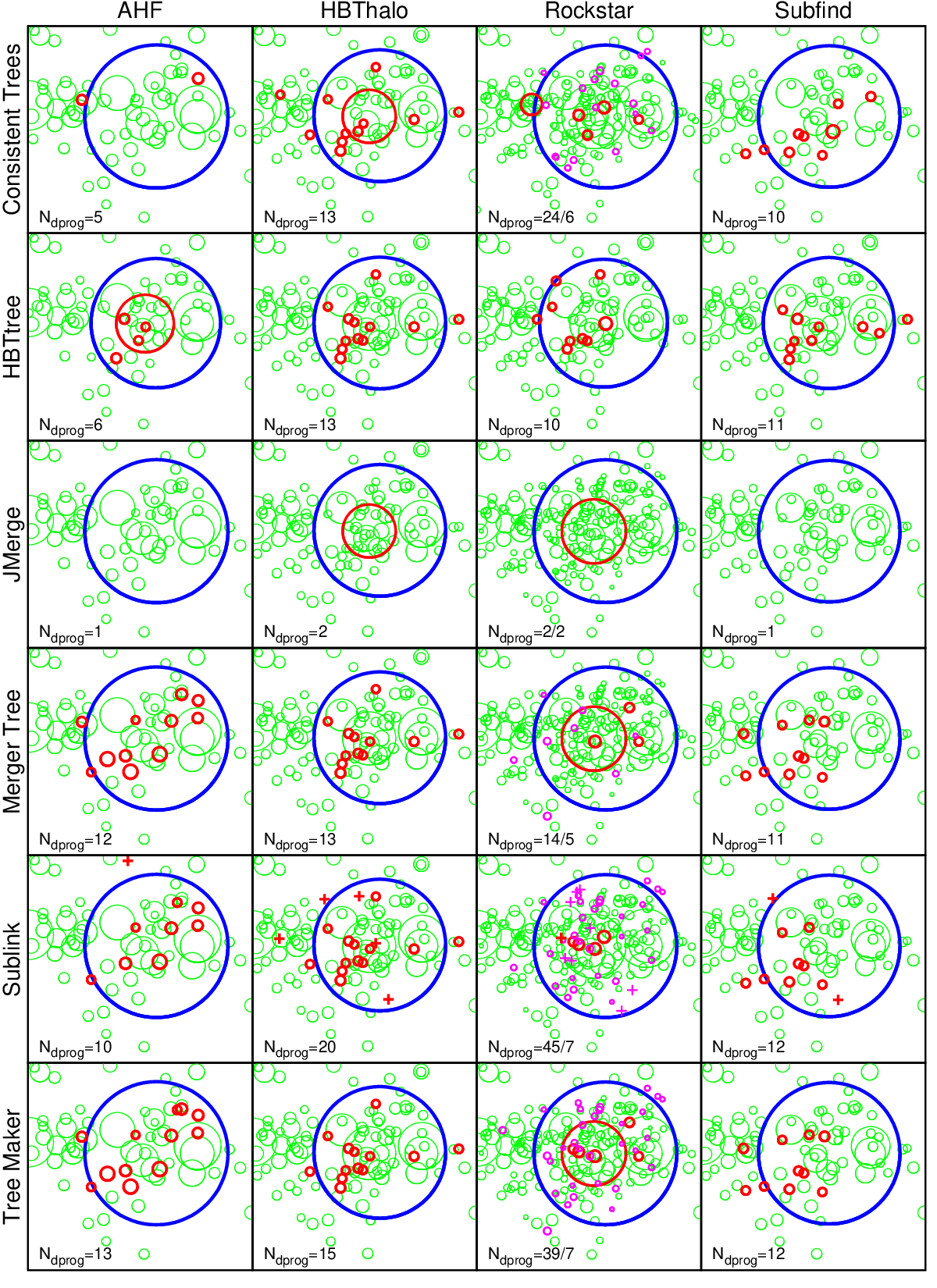}
  \caption{Projected image of a $3$ Mpc/h-side cube from snapshot 049
    centred on one of the most massive objects
    ($M>10^{14}h^{-1}M_{\sun}$) for all the combinations of halo
    finder (column) and tree builder (row). \velociraptor\ gives the
    same results as \treemaker, so it was omitted.  All blue (thickest line), red (intermediate-thick line), and
    magenta (intermediate-thin line) circles (of size $R_{200c}$) together with crosses
    represent \halos\ that will merge into the same halo in snapshot
    050. The blue halo is the main (and the biggest) progenitor, red and magenta circles
    are the remainder of direct progenitors (at snapshot 049), crosses
    (only for \sublink) represent \halos\ found in snapshot 048 but
    not in snapshot 049 which also merge into the same halo at
    050. \Halos\ in magenta have mass below $20m_p$ (only possible for
    \rockstar), whereas red \halos\ have larger masses.  Green (thinest line) circles
    are other \halos\ in the volume displayed.  The label at the
    left-bottom corner of each cell indicates the number of
    progenitors found for that combination, some of which may be
    missing if they are not in the visible volume. For \rockstar\ we
    show two numbers, the first one lists all \halos, the second when
    only those with mass larger than $20 m_p$ are considered. }
\label{case2}
\end{figure*}

Another simple tree property, which is nevertheless very important for
characterising the structure or geometry of a tree, is the number of
direct progenitors $N_{\rm dprog}$ (or local branches) that a halo
typically has. \Fig{fig:Nprog} shows the normalised (divided by the
total number of events) histogram of $N_{\rm dprog}$ for all \halos\
in the range $0\le z \le2$.  For all the various combinations of tree
building method and halo finder the most common situation is to have
just one single progenitor, corresponding to a halo having no mergers
on this step (which can happen multiple times during a \halos\
lifetime). The second most common situation is for a halo to have no
progenitors, which corresponds to a halo passing above the detection
threshold and appearing for the first time, which can happen only
once.  As for other properties studied in this paper, our results
would certainly change if we were to use a different set of output
times, so the importance does not lie in the individual tree results,
but in their differences.  For an elaborate study of the optimal
choice for the temporal spacing of snapshots to construct merger trees
we refer the reader to our upcoming paper (Wang Y. et al., in prep.)
or to past studies on the topic \citep{Benson12}.

It is noticeable that the \rockstar\ catalogue (blue dotted line)
yields a tree with significantly large branching ratio for the tree
builders \sublink, \treemaker, and \velociraptor. Also, besides using a very 
similar technique, \mergertree\ shows a more moderate branching ratio. 
By removing objects
with mass lower than $20 m_p$ (cyan dash-dotted line), we verified
that this high branching ratio is related to objects with very low
mass as these high-$N_{\rm dprog}$ cases disappear. Recall that, even
though all the halo finders cut their catalogue at 20 particles, for
\rockstar\ the mass $M_{200c}$ can be lower if some of those particles lay
outside $R_{200c}$. This small change, in general, moves the curves
for \rockstar\ from the highest branching ratio to the lowest one.
Note that the mass limited tree shown in cyan is not equivalent to the
other trees because the catalogue was reduced \textit{after} running
the tree building algorithm on it, hence giving non-self-consistent
trees.  Nevertheless, we do not expect great variations in
\Fig{fig:Nprog} between the cyan line and a fully self-consistent tree
with the same mass limit.  This serves as an illustration of the great
influence of the lower mass limit, pointing out again the importance
of the input halo catalogue in the resulting tree construction.

To illustrate a high branching ratio case we have selected one of the
extreme cases with $N_{\rm dprog}>30$ in \Fig{case2}. It corresponds to
one of the two most massive \halos\ (depending on the halo finder) at
snapshot 050 (z=0.32).  \Fig{case2} shows all the direct progenitors
of that halo and other \halos\ found in the area.  The blue (thickest) halo is
the main and most massive progenitor in the plot.  The red (intermediate-thick) and magenta (intermediate-thin)
circles represent other direct progenitors at snapshot 049 while green (thinest)
circles represent other (sub)\halos\ detected in the same region.
Magenta is used for \halos\ whose mass is below $20 m_p$ (only
possible for \rockstar), while red \halos\ have larger mass.
\sublink\ also has \halos\ that were found at snapshot 048, but were
not linked in snapshot 049, which were linked to the big halo at
snapshot 050; these are marked as crosses.

\Fig{case2} tells us that, when comparing different halo catalogues,
$N_{\rm dprog}$ tends to be correlated to the number of (small) \halos\
available to be absorbed, i.e. the more green \halos\ we find the more
merging (red and magenta) \halos\ we find.  We further confirm that
most secondary progenitors (red and magenta circles) are \subhalos\ of
the main progenitor (blue circle) and lie within $R_{200c}$. However,
in some cases secondary progenitors were found outside the volume
displayed (e.g. the halo missing in \consistenttree\ with \ahf).  But
in general, the properties of these \halos\ fit into the standard
merging picture in which \halos\ approaching a bigger one become
satellites (\subhalos), lose mass via tidal stripping and are
eventually totally absorbed.

If all the available \halos\ are considered, \rockstar\ is the
catalogue with most small \halos, leading to a higher branching ratio,
which drops when removing the low mass \halos. \hbthalo\ is also able
to discern more substructure, yielding a slightly higher $N_{\rm dprog}$
than \subfind\ and \ahf.

From the tree building point of view we remark that \sublink, with the
possibility of omitting one snapshot, increases $N_{\rm dprog}$
considerably for the two catalogues with more substructure: \rockstar\
and \hbthalo. \hbttree, in modifying the catalogue, tends to recover
the halo set generated by \hbthalo. This effect is more noticeable in
the case of \subfind\ because it is also based on FoF catalogues
(\Sec{sec:halos}). \jmerger\ shows very little branching ($N_{\rm dprog}=
1$ or $2$) because by construction it never associates a small merging
halo with a much bigger one. It rather associates the infalling halo
with another small halo.

Note, however, that this was a very extreme case and that \Fig{case2}
is not necessarily representative of the statistics seen in
\Fig{fig:Nprog}, rather it helps to understand the kind of factors
that influence the branching ratio.

\section{Mass Evolution} \label{sec:mass} 

The mass evolution of \halos\ is an important input for
semi-analytical models of galaxy formation.  In this section we will
study it through mass growth (\Sec{sec:growth}) and
fluctuations in mass (\Sec{sec:fluct}).

\subsection{Mass Growth} \label{sec:growth} 

Mass growth can be characterised by the discretised logarithmic
growth, defined as:

\begin{equation} \label{eq:alpha}
\frac{\rm{d}\log\ M}{\rm{d}\log\ t}\approx \alpha_M (k,k+1) = \frac{(t_k+t_{k+1})(M_{k+1}-M_k)}{(t_{k+1}-t_k)(M_{k+1}+M_k)}
\end{equation}

\noindent
where $k$ and $k+1$ are a halo and its descendant, with masses $M_k$
and $M_{k+1}$ at times $t_k$ and $t_{k+1}$, respectively
\citep{Srisawat13}.  In order to reduce the range of possible values of this
variable to the finite interval $(-1,+1)$, we define:

\begin{equation} \label{eq:beta}
\beta_M=\frac{1}{\pi/2}\rm{arctan}(\alpha_M)
\end{equation}

\begin{figure*}
  \centering
 \includegraphics[height=0.63\linewidth,angle=270]{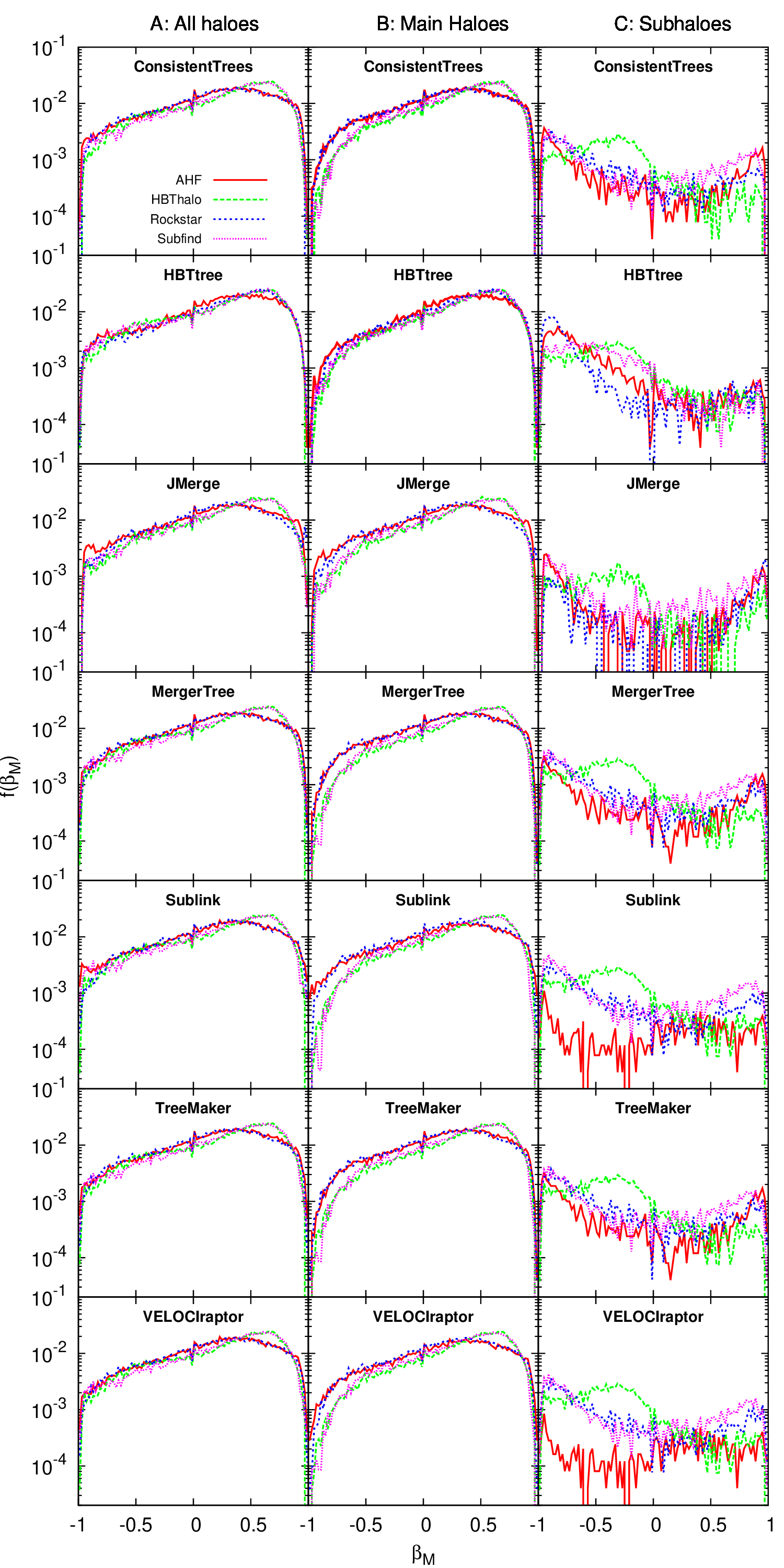}
 \caption{Mass growth distribution between two snapshots, $\beta_M$,
   related to the logarithmic mass growth through \Eq{eq:beta}, for
   \halos\ that can be identified at $z=0$, with mass $M>M_{th}$ at
   both output times.  We distinguish 3 populations: $A$ which
   contains all \halos\ with $M_{th}=M^{main}_{th}$, $B$ with only
   main \halos\ and $M_{th}=M^{main}_{th}$, and $C$ with only
   \subhalos\ and $M_{th}=M^{sub}_{th}$.  $M_{th}$ is tabulated in
   \Tab{tab:masscuts} for the different halo finders.  Each row
   displays a different tree building algorithm (as indicated). Each
   halo finder has its own line style as indicated in the legend.  The
   distribution is computed as a histogram, normalised by the total
   number of events found by the corresponding halo finder for the
   population $A$.  }
\label{fig:alpha}
\end{figure*}

\Fig{fig:alpha} shows the distribution of $\beta_M$ for three
populations: all \halos\ ($A$, on the left), main \halos\ ($B$, in the
centre) and \subhalos\ ($C$, on the right). All distributions have
been normalised by the total number of events found in halo sample $A$
in each case.  Selection is done as follows: all the \halos\
identified at $z=0$ are traced back along the main branch 
and at any snapshot if both a halo
and its descendant are main [sub] \halos\ and have mass $M>M_{th}^{\rm
  main}$ [$M>M_{th}^{\rm sub}$] (\Tab{tab:masscuts}) sum to the
population $B$ [$C$].  The population $A$ is compiled similarly, but
taking all pairs of \halos\ satisfying $M>M_{th}^{\rm main}$,
regardless of being main or \subhalos.  Note that the distribution $A$
is dominated by main \halos, since they are 
more numerous.

Within the hierarchical structure formation scenario one expects
\halos\ to grow over time.  This can be appreciated in column $A$,
where the distribution of $\beta_M$ is skewed towards values
$\beta_M>0$.  However, there is a non-negligible number of cases
($\sim 15-30\%$) where it decreases ($\beta_M<0$).  While mass loss
could be associated with tidal stripping of \subhalos, column $B$
shows that this is not the sole explanation within this simulation:
while \subhalos\ have an important contribution at the very far end of
the distribution (corresponding to large mass losses), there are also
many instances leading to $\beta_M<0$ for main \halos.  Nevertheless,
there are physical ways for main \halos\ to lose mass: when two main
\halos\ approach each other, the effective radius for tidal stripping
extends beyond the virial radius of the larger halo \citep[see][for an
elaborate discussion of exactly this phenomenon]{Behroozi13merger},
thus, the small one can experience mass loss before becoming a
satellite. Also, when \halos\ change their shape, the specific halo
mass definition (e.g. $M_{200c}$ for \ahf/\rockstar) of a halo finder
can lead to an apparent mass loss.

The plot clearly shows that the differences across halo finders are
greater than the variations introduced by the tree building method, with the
exception of \hbttree\ (that modifies the input halo catalogue).  There
are two distinct classes of distribution for main \halos\ ($B$): on
the one hand, \rockstar\ and \ahf, and on the other hand, \subfind\
and \hbthalo\ which have a more skewed distribution.  Recall from
(\Sec{sec:halos}) that the former use an inclusive mass definition,
thus, for a subhalo that just crossed the centre and is moving away,
the total (inclusive) mass of the host halo can decrease if part of
that subhalo crosses $R_{200c}$.

We finally remark that while \subhalos\ are present in our somewhat
low-resolution simulation (when compared to the state-of-the-art), they
contribute significantly to neither the shape nor the amplitude of the
mass growth distribution shown in column $A$ (all \halos). However,
their own distribution (column $C$) is interesting in its own regard:
we primarily observe mass loss due to tidal stripping, i.e. an
imbalance of the distribution towards negative $\beta_M$ values.  In
this case we find that whereas \hbthalo\ follows one distribution, the
other three follow their own. This reflects the inconsistency in
subhalo mass functions already seen in \Fig{fig:massfunc}.

In conclusion, most of the differences in the mass growth $\beta_M$
can be accounted for by the choices made by the respective halo finder
when defining quantities. In particular, \hbthalo\ and \subfind\ agree
best with the \textit{a priori} expectation from hierarchical
structure formation.

\subsection{Mass Fluctuations} \label{sec:fluct} 

After studying mass growth above, we quantify mass fluctuations by using

\begin{equation}
\xi_M = \frac{\beta_M(k,k+1)-\beta_M(k-1,k)}{2}
\label{eq:xi}
\end{equation}

\noindent
where ${k-1,k,k+1}$ represent consecutive timesteps. When far from
zero, it implies a growth followed by a dip in mass ($\xi_M<0$) or
vice versa ($\xi_M>0$).  Within the hierarchical structure formation
scenario this behaviour can be considered unphysical and equates to a
snapshot where the halo finder might not have assigned the correct
mass -- though there are certainly situations where the definition of
\textit{correct mass} remains arguable. Nevertheless, it provides
another means of quantifying the influence of the halo finder upon a
merger tree.

\begin{figure*}
\centering
\includegraphics[height=0.63\linewidth,angle=270]{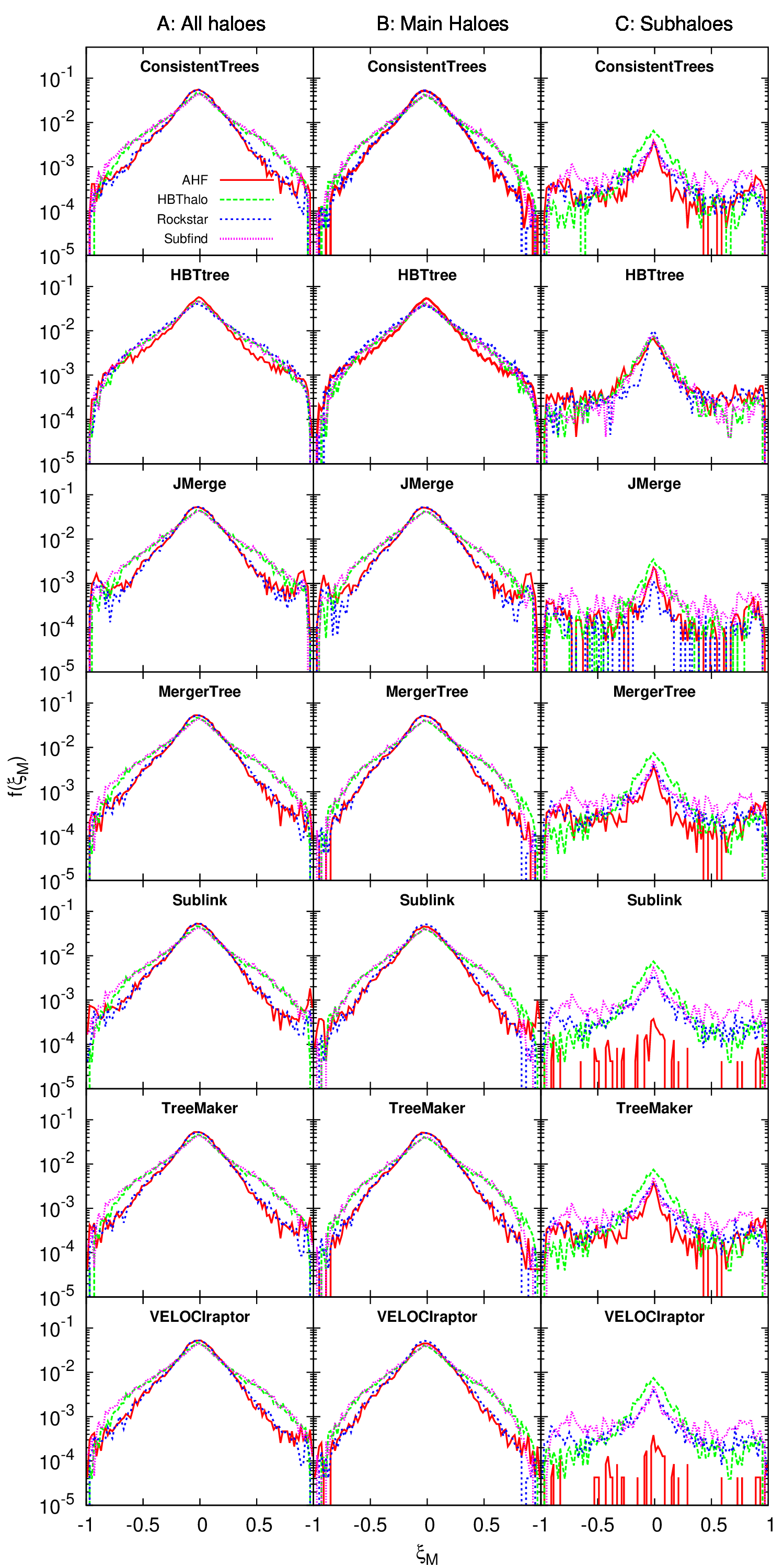}
\caption{Distribution of mass fluctuations $\xi_M$ (\Eq{eq:xi}), for
  \halos\ found in three consecutive snapshots along a main branch
  that can be identified at $z=0$, with mass $M>M_{th}$ for each
  appearance of the halo. We distinguish 3 populations: $A$ which
  contains all \halos\ with $M_{th}=M^{main}_{th}$, $B$ with only main
  \halos\ and $M_{th}=M^{main}_{th}$, and $C$ with only \subhalos\ and
  $M_{th}=M^{sub}_{th}$.  $M_{th}$ is tabulated in \Tab{tab:masscuts}.
  Comparison is made between different tree builders (each row as
  labeled) and halo finders (line styles as in the legend).  The
  distribution is computed as a histogram normalised by the total
  number of events for the corresponding halo finder for the
  population $A$.}
 \label{fig:xi}
\end{figure*}

The (normalised) distribution of $\xi_M$ is presented in \Fig{fig:xi}
in the same way as \Fig{fig:alpha}, i.e. three distinct columns for
all \halos\ ($A$, left), main \halos\ ($B$, middle), and \subhalos\
($C$, right).  It reconfirms most of the claims of
\Sec{sec:growth}. We again find the distribution is essentially
independent of the tree builder (besides \hbttree) for all three
populations.  We find two types of distributions for main \halos\
($B$): on the one hand, the \subfind\ and \hbthalo\ catalogues give
the broadest distributions and on the other hand, \rockstar\ and \ahf\
have a more peaked distribution. This implies that the first pair of
halo finders present more mass fluctuations ($\xi_M\neq0$) than the
second one. Note that this pairing is identical to the one reported in
\Sec{sec:growth}.  And we also find (again) that \subhalos\ ($C$) do
not provide an explanation for the wings of the mass fluctuation
distribution in column $A$, even though their own plot indicates that
they predominantly undergo abrupt changes, i.e. they have easily
distinguished wings.
 
\begin{figure}
\includegraphics[height=\figwidth,angle=270]{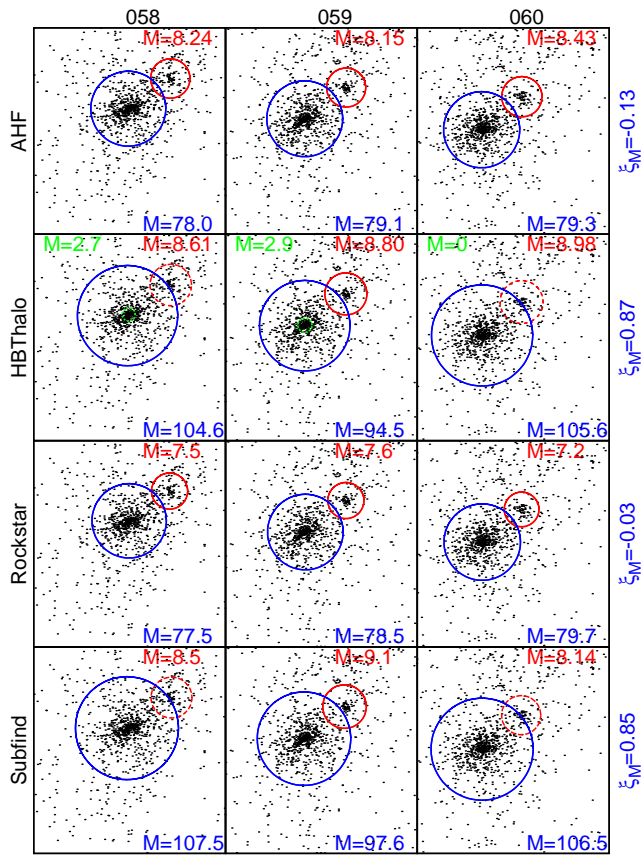}
\caption{Projected $1$ Mpc/h-side cube containing two \halos\ (three for
  \hbthalo) evolving from snapshot 058 (left column) to 059 (central
  column) to 060 (right column).  Each row shows a different halo
  finder. The radius of the circle is represented proportional to the
  mass of the object, with an extra factor of $\times 5$ for the small
  (red and green) \halos.  Dashed lines denote \subhalos\ whereas
  solid lines are used for main \halos.  The mass of each halo is also
  shown in units of $10^{10} h^{-1}M_{\sun}$.  
  At the right of each row we can see the value of
  $\xi_M$ for the big halo, which quantifies the mass fluctuation as
  defined by \Eq{eq:xi}.}
 \label{fig:case9}
\end{figure}

Given that \subhalos\ often undergo fluctuations (column $C$ of
\Fig{fig:xi}), this could cause fluctuations in main \halos\ when the
mass is defined exclusively (\hbthalo\ and \subfind).  In order to
study this effect, we selected a halo whose mass evolution is
characterised by a large $\xi_M$ value (for the \subfind/\hbthalo\ 
pair) in \Fig{fig:case9}.  We localised the same object (the big blue
halo) and surrounding ones (a red halo next to it and a green halo 
for \hbt\ in the center) in all four halo
catalogues, showing the three consecutive snapshots used for the
calculation of $\xi_M$ given at the very right hand side of each
panel. The halo undergoes a mass fluctuation for the finders \hbthalo\ 
and \subfind, while it keeps growing for \ahf\ and \rockstar.
\Fig{fig:case9} shows that, although it is true that for \hbthalo/
\subfind\ the total mass of the \subhalos\ increases when
the main halo decreases and vice versa, the fluctuation of subhalo
mass is one order of magnitude smaller than the main halo fluctuation
and this cannot be the sole explanation.  The fact that the red halo
changes from being a subhalo to a main halo and then back to a subhalo
again may be related (in a non-trivial way, since masses are defined
exclusively) to the mass fluctuation.
For this simple (compared to \Fig{case} \& \Fig{case2}) configuration of haloes,
all the tree building algorithms agree in the resulting trees. 
We also note that even small
fluctuations (10\% in mass) are detected by this parameter $\xi_M$, in
part due to an enhancement of $\xi_M$ at late times (cf. \Eq{eq:beta} \& \Eq{eq:xi}).

\subsection{Combining growth and fluctuations} \label{sec:massconclusion}

\begin{figure}
 
\includegraphics[height=\figwidth,angle=270]{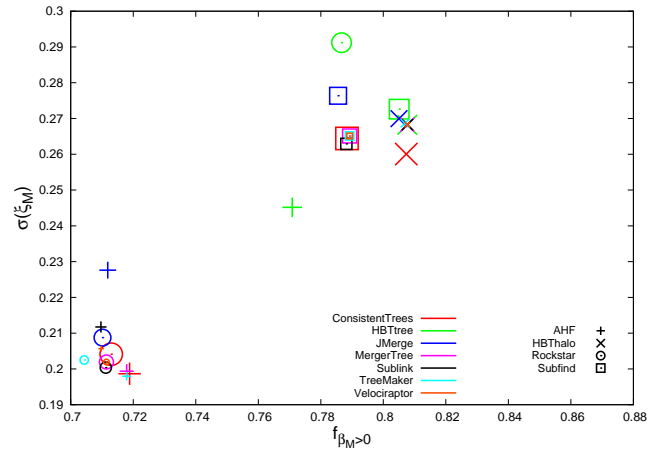}
\caption{Summary of \Fig{fig:alpha} and \Fig{fig:xi}. On the abscissa
  we show the fraction of \halos\ for which mass grows; on the
  ordinate we show the standard deviation of the mass fluctuations.
  Only main \halos\ satisfying $M>M_{th}^{main}$ (\Tab{tab:masscuts})
  are taken into account.  Every point represents a combination of a
  tree builder (size and colour-coded) and a halo catalogue
  (symbol-coded, see legend).  
  The size of the points represents the different tree builders in decreasing 
  order as they are listed in the legend (\consistenttree\ the largest, 
  \velociraptor\ the smallest).
  }
\label{summary}
\end{figure}

To better draw any conclusion from our study of the mass evolution of
\textit{main \halos} we summarise results from their $\beta_M$ and
$\xi_M$ statistics (\Sec{sec:growth} \& \Sec{sec:fluct}) in
\Fig{summary}: the $x$-axis shows the fraction $f_{\beta_M>0}$ of
objects for which $\beta_M>0$, whereas the $y$-axis shows the standard
deviation $\sigma_{\xi_M}$ of $\xi_M$.  Different sizes (or colours) now
represent different tree building methods whereas the symbols stand
for the input halo catalogue.  The desirable feature of a tree
describing hierarchical structure formation would be to have small
mass loss for main \halos\ (high $f_{\beta_M>0}$) and small mass
fluctuations (low $\sigma_{\xi_M}$), at least \textit{a priori},
because we also explained physical causes for these phenomena. Note
also that the quantities plotted here do not provide a substitute for
the whole curve shown in \Fig{fig:alpha} \& \Fig{fig:xi}, but rather
capture well the features of interest as they are observed.  This
summary plot illustrates very well how mass evolution sensitively
depends on the choice of the halo finder:

\begin{itemize}
\item Points for the same halo finder (symbol) group together. The
  small scatter amongst those groups represents the small influence of
  the tree building method on these magnitudes.
\item \hbttree\ points deviate from the group, approaching the
  area of the \hbthalo\ finder (crosses).
\item The pair of halo finders \hbthalo/\subfind\ achieves a lower rate of
  mass loss at the price of having more mass fluctuation than the
  other pair of finders \ahf/\rockstar\ for main \halos. We relate
  this pairing to the mass definition of the halo finder: the former
  is exclusive and uses self-bound objects, whereas the latter uses
  inclusive spherical $M_{200c}$ objects.
\end{itemize}

We have verified that mass growth and fluctuations are intrinsically
related to the mass definition.  A simple change from an inclusive to
an exclusive halo catalogue or from $M_{200c}$ to arbitrarily shaped
\halos\ would change the shape of the curves seen in \Fig{fig:alpha}
\& \Fig{fig:xi} and the position of the points in \Fig{summary}.  But
other fundamental properties of the halo finder also leave their
imprint, the evident differences between \hbthalo\ and \subfind\ in
\Fig{summary} are a proof of this.

\section{Discussion \& Conclusions} \label{sec:disc}

Following the first paper \citep[i.e.][]{Srisawat13} in a series of
articles comparing various tree building codes, we investigated the
influence of the input halo catalogue on the quality of the resulting
merger trees. 'Quality' in this regard has been identified as length
of the main branch, number of direct progenitors, and, quantities that
are highly relevant for semi-analytical modeling, the mass growth and
mass fluctuation of \halos.  We also showed some specific examples of
cases that aided our understanding of the influence of the halo finder
and tree builder on the resulting properties of the trees.

In total, seven different tree building methods have been applied to
the halo catalogues produced by four different halo finding algorithms
which examined the same cosmological simulation. This produced 28
merger trees to be analysed. The influence of both groups of codes is
summarised below, and the particular achievements and difficulties of
the different methods discussed.

\subsection*{The influence of the halo finder}

The primary conclusion of all the studies presented here is that the
influence of the input halo catalogue is greater than the influence of
the tree building method employed.  This is especially clear for the
mass evolution studies (\Sec{sec:mass}) although it is also noticeable
from the results of the main branch length (\Sec{sec:length}) and the
studies on the branching ratio also suggest it (\Sec{sec:branches}).
Part of these differences are due to the fact that for this comparison
we allowed the halo finders to choose their own definitions instead of
unifying them as done in previous halo finder comparison
projects. However, this way we find the real impact a user will
encounter when choosing one or the other halo finder for his/her
analysis.

Another pattern encountered along our studies is the pairing
\ahf/\rockstar\ vs. \hbthalo/\subfind. This is very clear in the mass
evolution of main \halos\ (central columns of \Fig{fig:alpha} and
\Fig{fig:xi}, summarised in \Fig{summary}) and can also be seen in the
main branch length distribution (\Fig{length}).  We interpret this
pairing to be caused by the fundamental construction of the halo
catalogues, namely spherically truncated $M_{200c}$ inclusive masses
(\Eq{eq:virialradius}) for the former pair vs. self-bound exclusive
objects starting from FoF groups for the latter. These differences can
already be acknowledged in the main halo mass function shown in
\Fig{fig:massfunc}.

The studies on the length of the tree (\Sec{sec:length}) are the
cleanest test, since they do not rely on arbitrary choices such as the
lower mass cut (which makes a significant difference for the branching
ratio) or the mass definition (which is of great influence in the mass
evolution). The tracking nature of \hbthalo\ showed excellent results
in this section, with no early truncation of (sub)\halos.  \rockstar\
and \ahf\ showed early truncation of trees, especially for \subhalos\
near the centre of their host, whereas \subfind\ did not show too much
early truncation of \subhalos, because they are systematically missing
in the centre of the hosts.  \ahf, with its poor completeness at the
low-mass end led to the shortest main branches: because \halos\
disappear due to this incompleteness the main branches tend to end
early.

The relevance of the lower mass cut was also seen in the study of the
branching ratio (\Fig{fig:Nprog} in \Sec{sec:branches}). In
particular, for \rockstar\ a cut in mass was not equivalent to a cut
in the number of particles.  Because of this, doing the same cut in
particles as for other catalogues, the branching ratio of \rockstar\
was too high.

The mass evolution of \halos\ was found to be mostly dependent upon
the mass definition employed by the halo finder.  However, it is not
clear which finders perform best: \hbthalo/\subfind\ show less mass
loss whereas \ahf/\rockstar\ show fewer mass fluctuations.  Mass
evolution is intrinsically related to the way the mass is defined, and
the choice of a different mass definition within the same halo finder
would lead to different results.

Along these lines, note that some properties of the halo finders are
simple choices that are relatively easy to change, as for example the
exclusive/inclusive mass assignment or the choice of spherical \halos\
vs. self-bound objects.  However, we have seen in \citet{Knebe13} that
other, more fundamental, details of each halo finder (such as the
initial particle collection) leave their own unique signature in the
catalogue. These are practically unavoidable and hence the user has to
decide upfront which halo finder best suits
their needs.

\subsection*{The influence of the tree building method}

Although we found a greater dependence on the halo finder than on the
tree building method, each of the tree codes also has its own
peculiarities:

\begin{itemize}
\item \consistenttree\ in many cases is able to correct the problems
  posed by the finder by adding artificial \halos.
\item \hbttree, when recomputing the substructure, makes \halos\ more
  trackable, improving the results.
\item \jmerger\ has problems in dealing with the motion of (sub)haloes 
in highly clustered environments.
\item \mergertree, \treemaker\ and \velociraptor\ behave very
  similarly, as they are based on nearly identical algorithms.
\item \sublink\ is sometimes able to compensate for non-detection of
  \halos\ by looking at non-consecutive timesteps.
\end{itemize}

\subsection*{Outlook}

The main outcome of the present paper is that the fundamental
properties of halo finders have a major impact on the merger trees
constructed from them, and that some tree building techniques can help
improve those trees by correcting for halo finder defects.  We pointed
out the repercussions that several properties of the halo finders and
tree building codes can have on the final trees.  This should help the
community choosing, designing or modifying their pipelines to
construct merger trees idealised for their specific purposes.

It is worth mentioning that, although here we focused on the
differences among the resulting merger trees, the agreement among them
is nevertheless remarkable.  The general features of the trees
resulted as one would have expected, and are similar from one tree to
another. Many times the differences between trees are only seen when
plots are done on a logarithmic scale, since those differences are at
the order of a few-cases for every thousand plotted.

The remaining question is how all this affects our understanding of
the Universe, at least when theoretically modeling it.  This series
of code comparison workshops helps us exploring the degree of
certainty we have when generating virtual skies.  The first
workshop(s) related to the identification of objects: \halos,
\subhalos\ and galaxies. We are currently investigating the linkage
between objects: the merger trees.  In an ultimate step we will
analyse how the different pipelines can lead to different simulated
direct observables: from the merger trees we will move to the effect
of semi-analytic methods and other ways to generate galaxy mock
catalogues and placing them in lightcones to generate mock
surveys.\footnote{In the upcoming workshop: http://popia.ft.uam.es/nIFTyCosmology}

\section*{Acknowledgements} \label{sec:Acknowledgements}

The {\sc Sussing Merger Trees} Workshop was supported by the European
Commission's Framework Programme 7, through the Marie Curie Initial
Training Network CosmoComp (PITN-GA-2009-238356).  This also provided
fellowship support for AS.

SA is supported by a PhD fellowship from the Universidad Aut\'onoma de Madrid
and the Spanish ministerial grants AYA2009-13936-C06-06, AYA2009-13936-C06-06, FPA2012-39684-C03-02 and SEV-2012-0249.
He also acknowledges the support of the University of Western Australia through their Research Collaboration Award 2014 
scheme and thanks its International Centre for Radio Astronomy Research 
(and especially Chris Power) for the hospitality during the final stages of the paper writing.

AK is supported by the {\it Ministerio de Econom\'ia y Competitividad} 
(MINECO) in Spain through grant AYA2012-31101 as well as the 
Consolider-Ingenio 2010 Programme of the {\it Spanish Ministerio 
de Ciencia e Innovaci\'on} (MICINN) under grant MultiDark CSD2009-00064. 
He also acknowledges support from the {\it Australian Research Council} 
(ARC) grants DP130100117 and DP140100198. 
He further thanks Belle \& Sebastian for tigermilk.

PSB is funded by a Giacconi Fellowship through the Space Telescope Science 
Institute, which is operated by the Association of Universities for Research 
in Astronomy, Incorporated, under NASA contract NAS5-26555.

PJE is supported by the SSimPL programme and the Sydney Institute for
Astronomy (SIfA).  

JXH is supported by an STFC Rolling Grant to the Institute for
Computational Cosmology, Durham University.

CS is supported by The Development and Promotion of Science and Technology 
Talents Project (DPST), Thailand.

PAT acknowledges support from the Science and Technology Facilities
Council (grant number ST/I000976/1).

YYM received support from the Weiland Family Stanford Graduate Fellowship.

The authors contributed in the following ways to this paper: PAT, FRP,
AK, CS, AS, organised the workshop from which this project
originated. They designed the comparison, planned and organised the
data. The analysis presented here was performed by SA (a PhD student
of AK) and the paper was written by SA and AK. The other authors
provided results and descriptions of their respective algorithms; they
also contributed towards the content of the paper and helped to
proof-read it.

\bibliography{mn-jour,treebuild}
\bibliographystyle{mn2e} \label{sec:Bibliography}

\label{lastpage}
\
\end{document}